\documentclass[lettersize,journal]{IEEEtran}
\usepackage{amsmath,amsfonts,amssymb}
\usepackage{algorithm,algpseudocode}
\algnewcommand{\algorithmicforeach}{\textbf{for each}}
\algdef{SE}[FOR]{ForEach}{EndForEach}[1]
  {\algorithmicforeach\ #1\ \algorithmicdo}
  {\algorithmicend\ \algorithmicforeach}
\usepackage{enumerate}
\usepackage{algpseudocode}
\usepackage{array}
\newcolumntype{?}{!{\vrule width 1pt}}

\usepackage[caption=false, labelformat=parens, subrefformat=parens]{subfig}
\usepackage{textcomp}
\usepackage{stfloats}
\usepackage{url}
\usepackage{diagbox}
\usepackage{verbatim}
\usepackage{graphicx}
\usepackage{cite}
\usepackage{booktabs}
\usepackage[hidelinks]{hyperref} 
\usepackage{makecell}
\usepackage{xcolor}
\usepackage[export]{adjustbox}
\usepackage{array, multirow}
\usepackage{tabularx}
\usepackage[switch]{lineno}
\usepackage{pifont}

\begin{document}
\title{From High-SNR Radar Signal to ECG: A Transfer Learning Model with Cardio-Focusing Algorithm for Scenarios with Limited Data}

\author{Yuanyuan~Zhang, Haocheng Zhao, Sijie Xiong, Rui~Yang, \IEEEmembership{Senior~Member,~IEEE}, \\ Eng~Gee~Lim, \IEEEmembership{Senior~Member,~IEEE,  Yutao~Yue, \IEEEmembership{Senior~Member,~IEEE} 
\thanks{This research has been approved by the University Ethics Committee of Xi'an Jiaotong-Liverpool University with proposal number ER-SAT-0010000090020220906151929, and is partially supported by National Natural Science Foundation of China (72401233), Jiangsu Provincial Scientific Research Center of Applied Mathematics (BK20233002), Suzhou Science and Technology Programme (SYG202106), Jiangsu Industrial Technology Research Institute (JITRI) and Wuxi National Hi-Tech District (WND). \textit{(Corresponding authors: Rui Yang, Yutao Yue.)}}
\thanks{Yuanyuan Zhang and Haocheng Zhao are with the School of Advanced Technology, Xi'an Jiaotong-Liverpool University, Suzhou, 215123, China, the Department of Electrical Engineering and Electronics, University of Liverpool, Liverpool, L69 3GJ, United Kingdom, and also with the Institute of Deep Perception Technology, JITRI, Wuxi, 214000, China (email: Yuanyuan.Zhang16@student.xjtlu.edu.cn, Haocheng.Zhao19@student.xjtlu.edu.cn).}
\thanks{Sijie Xiong is with the Faculty of Information Science and Electrical Engineering, Kyushu University, 819-0395, Fukuoka, Japan (email: xiong.sijie.630@s.kyushu-u.ac.jp).}
\thanks{Rui Yang and Eng Gee Lim are with the School of Advanced Technology, Xi'an Jiaotong-Liverpool University, Suzhou, 215123, China (email: R.Yang@xjtlu.edu.cn; Enggee.Lim@xjtlu.edu.cn).}
\thanks{Yutao Yue is with the Thrust of Artificial Intelligence and Thrust of Intelligent Transportation, The Hong Kong University of Science and Technology (Guangzhou), Guangzhou 511400, China, and also with the Institute of Deep Perception Technology, JITRI, Wuxi 214000, China. (email: yutaoyue@hkust-gz.edu.cn).}}}

\maketitle

\begin{abstract}
Electrocardiogram (ECG), as a crucial fine-grained cardiac feature, has been successfully recovered from radar signals in the literature, but the performance heavily relies on the high-quality radar signal and numerous radar-ECG pairs for training, restricting the applications in new scenarios due to data scarcity. Therefore, this work focuses on radar-based ECG recovery in new scenarios with limited data and proposes a cardio-focusing and -tracking (CFT) algorithm to precisely track the cardiac location to ensure an efficient acquisition of high-quality radar signals. Furthermore, a transfer learning model (RFcardi) is proposed to extract cardio-related information from the radar signal without ECG ground truth based on the intrinsic sparsity of cardiac features, and only a few synchronous radar-ECG pairs are required to fine-tune the pre-trained model for ECG recovery. The experimental results reveal that the proposed CFT can dynamically identify the cardiac location, and the RFcardi model can effectively generate faithful ECG recoveries after using a small number of radar-ECG pairs for training. The code and dataset will be made available after publication.
\end{abstract}

\begin{IEEEkeywords}
Contactless Vital Sign Monitoring, Radar-Based Sensing, Transfer Learning, Derivative-Free Optimization
\end{IEEEkeywords}

\section{Introduction}
Radio detection and ranging (radar) system was originally designed for military detection of large aircraft by emitting electromagnetic waves and evaluating the reflections. The follow-up research has investigated the civilian use of radar systems for contactless sensing in various scenarios, such as autonomous driving~\cite{yao2023waterscenes} and human monitoring~\cite{zhang2023overview}. Over the past decade, radar-based sensing has been empowered by deep neural networks to process non-stationary reflected signals or high-dimensional data, enabling versatile applications to replace contact- or visual-based measurement for convenience or privacy concerns (e.g., vital sign monitoring~\cite{liu2024diversity}, gesture recognition~\cite{liu2025mhtrack}, person identification~\cite{wang2025open}).

Radar-based vital sign monitoring, as a popular branch of radar-based sensing, has been explored for decades to measure heart rate or respiration rate in a contactless manner~\cite{zhang2023overview}, and some further studies leverage the deep neural network to realize domain transformation from cardiac mechanical activities (i.e., heartbeat) to electrical activities (i.e., electrocardiogram (ECG)), providing a fine-grained cardiac measurement for wellness monitoring or clinical diagnosis~\cite{zhao2024mmarrhythmia,zhang2024radarODE,zhang2024radarODE-MTL,chen2022contactless,zhang2025horcrux,li2024radarnet,wu2023contactless}. \textcolor{black}{In the literature, radar-based ECG recovery is only realized by deep-learning-based methods, because the domain transformation is extremely complex to be modeled mathematically while such transformation can be learned by a deep learning model due to the great nonlinear mapping ability~\cite{chen2022contactless}.}

Similar to other research fields involved with deep learning, radar-based ECG recovery also asks for numerous radar signals to train the deep learning model with synchronous ECG ground truths~\cite{li2024radarnet,zhang2024radarODE-MTL,xiong2025enhancing}. According to previous research, the performance of the deep learning model degrades heavily after reducing $30\%$ of the training data even after applying proper data augmentation techniques~\cite{zhang2025horcrux}, causing difficulties for the deployment in new scenarios due to the demand for hours of ECG collection~\cite{zhang2024radarODE}. However, the method for reducing dependence on data quantity is rarely investigated for radar-based ECG recovery, and all the deep-learning-based ECG recovery models are trained in a supervised manner with large dataset containing $3-32$ hours of synchronous radar-ECG pairs~\cite{chen2022contactless,zhao2024airecg,li2024radarnet}.  

In contrast, most studies are dedicated to inventing advanced signal processing algorithms to enhance the signal quality, because the deep learning model for ECG recovery is vulnerable to the inputs contaminated by noise and requires high signal-to-noise ratio (SNR) radar signals as inputs~\cite{chen2022contactless,liu2024diversity}. The methods for capturing high-SNR radar signals can be categorized into two groups:
\begin{itemize}
\item The first type of method focuses on designing advanced radar front-end with multiple transmitters (Tx) and receivers (Rx)~\cite{li2024robust,xiong2022vital} or calibrating baseband radar signals from in-phase and quadrature (IQ) channels to a circular shape~\cite{dong2024robust,ni2024accurate,zhang2024single}.
\item The second type of method assumes that the rough localization of human body provides accurate chest region with the majority of range bins containing useful cardiac features, and high-SNR signal can be obtained by selecting useful channels~\cite{li2024radarnet,zhang2025umimo}, applying clustering algorithms~\cite{chen2022contactless} or accumulating the signals from various dimensions (e.g., chirps, frames, antennas)~\cite{liu2024diversity}. 
\end{itemize}

The first type of method is not suitable for some commonly used frequency-modulated continuous-wave (FMCW) radar platforms (e.g., TI AWR-x radar) due to the on-board digital front-end module filtering the frequency-modulated feature of baseband signal (i.e., circular IQ plot)~\cite{chen2024co}, preventing the broad applications of this approach in commercial radar. The second type of method relies on accurate localization of the chest region, while existing methods only provide a rough location of the human body, causing a deviation of several decimeters due to different postures of the subject~\cite{chen2021movi}. Therefore, the methods based on signal accumulation may fail because only a minority of range bins contain cardiac features, hence not subjecting to the law of large numbers~\cite{liu2024diversity}. Although some aforementioned studies have proposed methods for selecting or clustering the useful range bins with cardiac features~\cite{chen2022contactless,li2024radarnet}, the computational cost for traversing a large objective space can be huge without an accurate cardiac location.

Based on the above discussion, it is still a challenge to: (a) precisely locate and track the cardiac location during data collection to efficiently extract high-SNR radar signal; (b) develop a deep learning framework for radar-based ECG recovery with less demand for ECG collection to realize an efficient model training especially for new scenarios with limited data. To overcome these two challenges, the contributions of this study can be listed as: 
\begin{itemize}
\item A cardio-focusing and -tracking (CFT) algorithm is proposed based on derivative-free optimization (DFO) to find the cardio-focused (CF) point by iteratively evaluating the potential points in a discontinuous objective space, with a universal signal template designed to adaptively assess the signal SNR as costs.
\item A transfer learning framework RFcardi is proposed following a self-supervised learning (SSL) paradigm to effectively learn the latent representations from radar signals by leveraging an appropriate pre-text task. Accordingly, this work further designates the sparse signal recovery (SSR) as the pre-text task, assisting the RFcardi to learn essential representations for the later ECG recovery.
\item The proposed CFT algorithm has been validated on sitting subjects in various scenarios and could provide radar measurements with better SNR compared with existing methods. In addition, the pre-trained RFcardi framework can be easily adapted to realize radar-based ECG recovery with a small number of synchronous radar-ECG measurements for fine-tuning. 
\end{itemize}

The rest of the paper is organized as follows. Section~\ref{sec:rw} provides the background information for radar-based ECG recovery and SSL. Section~\ref{sec:method} elaborates the proposed CFT algorithm and RFcardi framework, with the experimental settings and results shown in Section~\ref{sec:exp} and~\ref{sec:result}. The final conclusion is shown in Section~\ref{sec:conclusions}.

\section{Theoretical Background and Challenges}\label{sec:rw}
\subsection{FMCW Radar Foundations}
FMCW radar has been widely used in nowadays millimeter-wave (mmWave) sensing to measure the range, velocity and angle of arrival (AoA) of the objects appearing in the field of view~\cite{tang2024bsense}, with three critical concepts that configure the transmitted waveform:
\begin{itemize}
\item \textbf{Chirp } is the minimum component in the FMCW signal with microsecond-level duration and is often called fast time. The waveform of a single chirp is a sinusoidal signal with frequency that changes linearly over time, with the key characteristics designated by start frequency, bandwidth and chirp duration to get range information of the object.
\item \textbf{Frame} is a collection of multiple chirps that forms a complete observation window to get the velocity information based on the range bins extracted from chirps and is often referred to as slow time.
\item \textbf{Virtual antenna array (channel)} is a commonly used concept in multiple-input and multiple-output (MIMO) radar systems and is able to realize complex modulations or beamforming~\cite{xiong2022vital}. However, this study mainly leverages the phase difference across antenna channels to estimate the AoA of the objects.
\end{itemize}
The popular commercial radar platforms have provided a convenient interface for radar configuration,  signal modulation and demodulation~\cite{AWR1843}. Therefore, detailed equations of FMCW radar signal processing might be redundant in this paper, while the theoretical explanation can be found in previous papers~\cite{wang2021mmhrv,tang2024bsense}.

\subsection{Cardiac Signal Extraction from FMCW Radar}
ECG recovery relies on the high-SNR radar inputs that describe the mechanical cardiac activities, and the reflected signal from a given point $E=(x,y,z)$ in 3D space can be expressed as:
\begin{equation}\label{equ:raw_sig}
R(E, t)=\sum_{v=1}^V \sum_{c=1}^C \sum_{n=1}^N s_{v, c, n}(t)\cdot e^{j 2 \pi \frac{2k\cdot d(E, v)}{\text{light speed}} n} \underbrace{e^{j 2 \pi \frac{2\cdot d(E, v)}{\lambda}}}_{\text{phase term}\ \phi}
\end{equation}
where $(x,y,z)$ represents the (horizontal, radial, vertical) axis, $V$ is the number of virtual antenna channels, $C$ means the number of chirps within one frame, $N$ is the total sample points within one chirp, $s_{v, c, n}(t)$ denotes the original received signal, $k$ is the slope of frequency raising, $\lambda$ means wavelength and $d(E, v)$ represents the distance between point $E$ and virtual antenna $v$~\cite{chen2022contactless}. In FMCW processing for vital sign monitoring, the time sample $t$ corresponds to one frame instead of the sample point $n$, and the signal from different chirps $c$ and antenna channels $v$ will be accumulated to improve SNR~\cite{liu2024diversity}.

The interested term in (\ref{equ:raw_sig}) is the variation of distance $d(E,v)$, because it represents the displacements caused by respiration and heartbeat (without considering any other noise). Therefore, the chest region displacement $h(E, t)$ can be unwrapped from phase variation $\Delta \phi$ as
\begin{equation}\label{equ:phase}
h(E, t) = \frac{\lambda \Delta \phi}{4\pi}
\end{equation}

\begin{figure}[tb]
        \centering
        \subfloat[]{\label{fig:radar_good}\includegraphics[width=0.5\columnwidth]{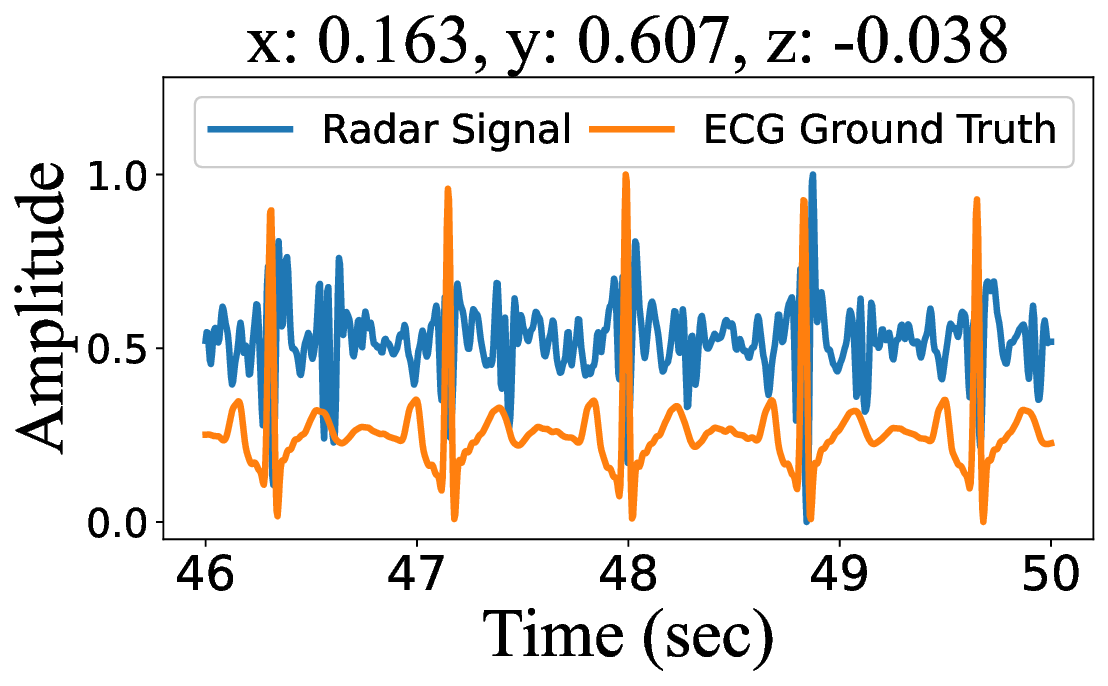}}
        \subfloat[]{\label{fig:radar_bad}\includegraphics[width=0.5\columnwidth]{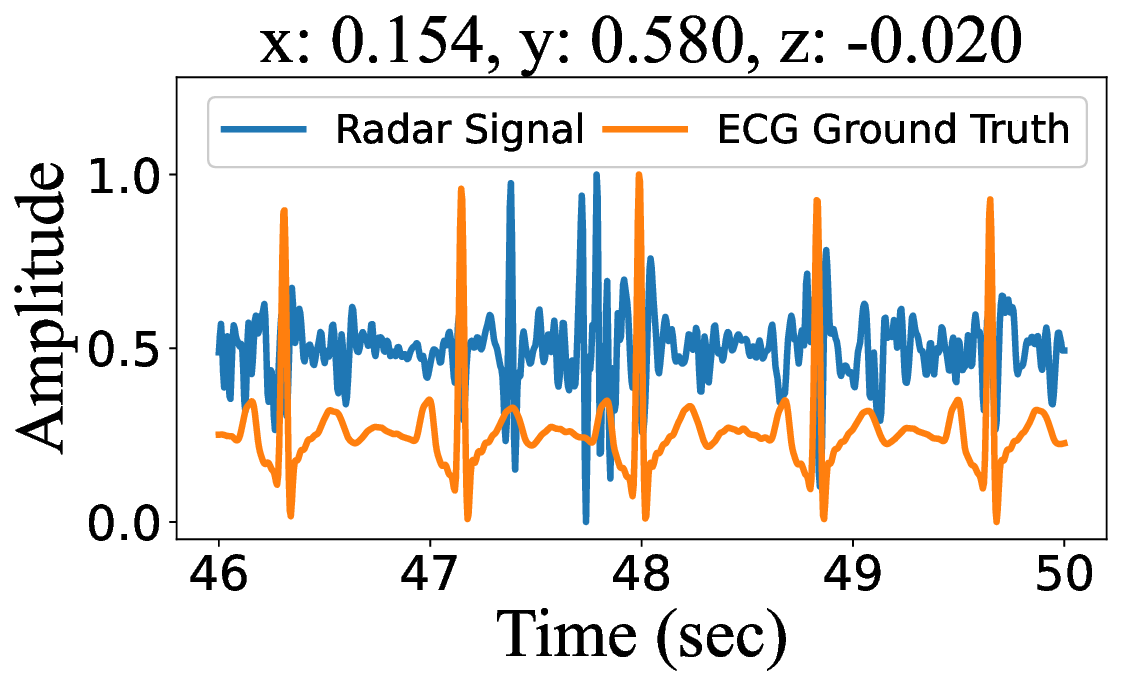}} \\
        \subfloat[]{\label{fig:org_scen}\includegraphics[width=0.5\columnwidth]{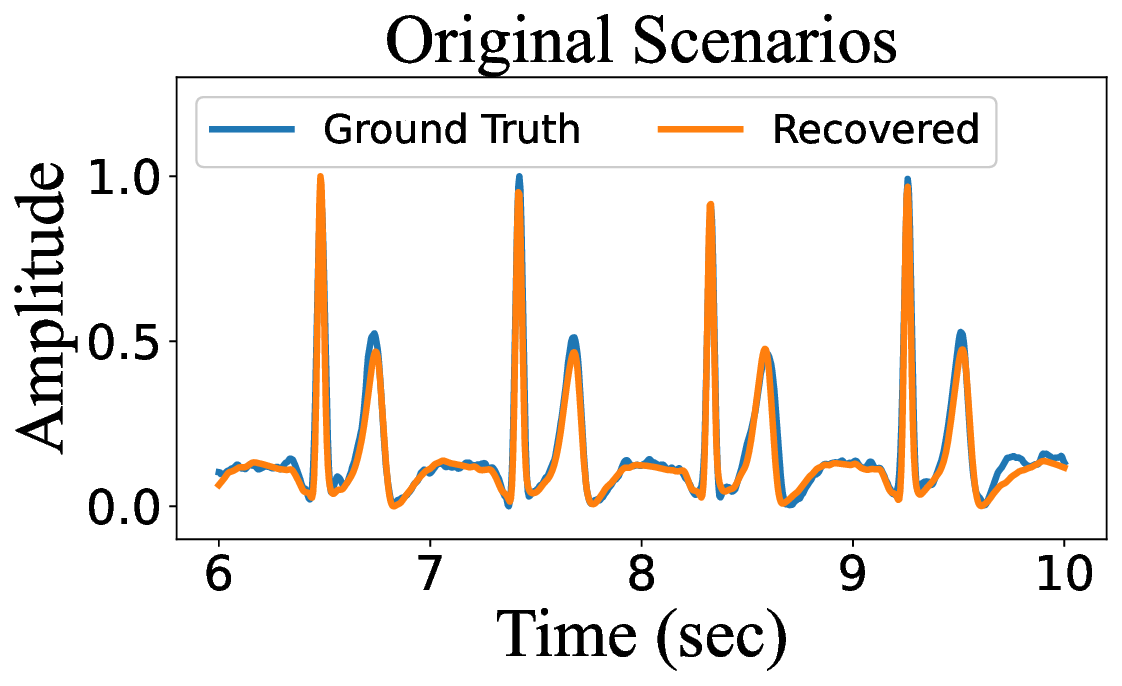}}
        \subfloat[]{\label{fig:new_scen}\includegraphics[width=0.5\columnwidth]{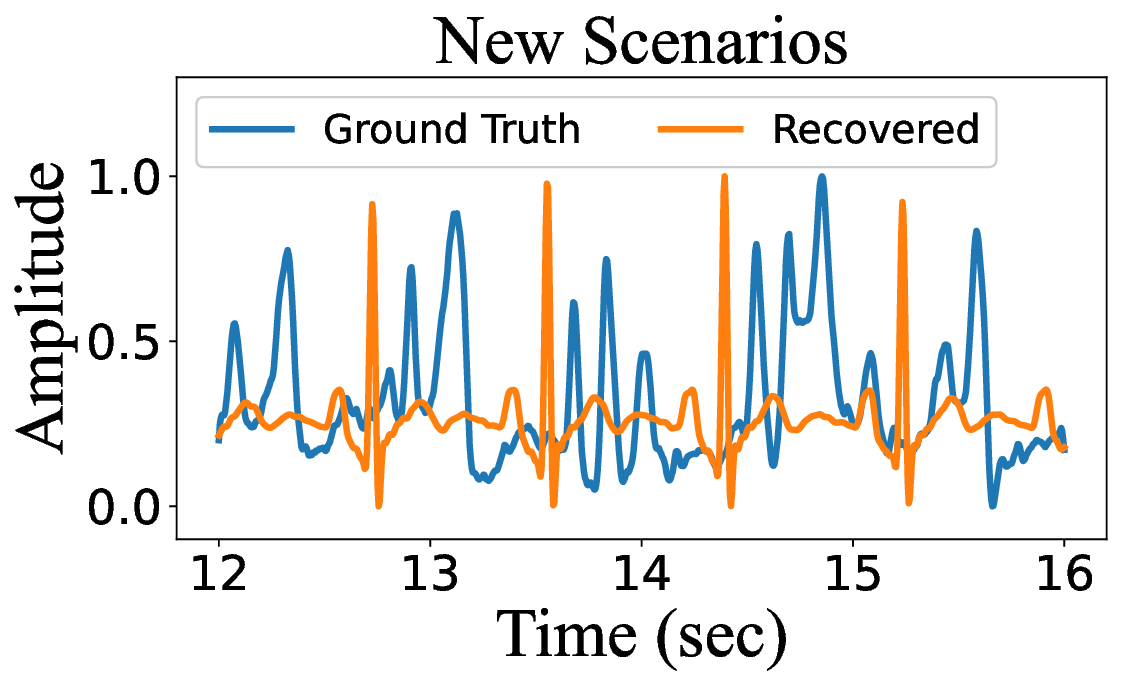}}
        \caption{\textcolor{black}{Challenges for radar-based ECG recovery: (a) and (b) Radar signals with high and low SNR for adjacent points with a distance of $0.03$m; (c) and (d) Inference results of a well-trained deep learning model in original and new scenarios.}}
        \label{fig:challenges}
\end{figure}

\color{black}
At last, some common noises, such as respiration and thermal noise, can be easily removed using a band-pass filter and differentiator to make sure that the final $h(E, t)$ mostly contains cardiac-related features from point $E$ as shown in Figure~\subref*{fig:radar_good}, with prominent vibration corresponding to the R peaks in the ECG signal.

\textbf{Challenge: } It is natural to think the high-SNR radar signal can be searched in a constrained space by optimization, while there is no appropriate method to assess the signal SNR in terms of cardiac features contained. In addition, the objective space is highly discontinuous, and the signals extracted from adjacent points reveal totally different SNR, as indicated by the poor radar collection with much noise that buries the cardiac features (peaks) in Figure~\subref*{fig:radar_bad}, restricting the application of common gradient-based optimization algorithms.
\color{black}

\subsection{Transfer Learning}
In addition to high-SNR radar inputs, accurate ECG recovery also relies on the scale of datasets to train the deep learning model, with the previous research normally adopting the dataset with the length of $3-32$ hours~\cite{chen2022contactless,zhao2024airecg,li2024radarnet}. Unfortunately, the initial experimental results show that the well-trained deep learning model cannot be directly used for the signal collected from new scenarios even using a similar radar configuration as shown in Figure~\subref*{fig:org_scen} and ~\subref*{fig:new_scen}, because different indoor scenarios and radar configurations may have unknown interference on radar signals~\cite{zhang2023overview}.

Inspired by other signal-based research~\cite{zhang2025umimo,chen2024tfpred}, transfer learning is a promising paradigm to learn the latent representation from unlabeled radar signal to capture basic cardiac features in an SSL manner, reducing the need for cumbersome ECG collection. Then, only a small number of synchronous radar-ECG pairs is required to fine-tune the pre-trained model to realize the ECG recovery for new scenarios.

\textbf{Challenge: } The efficient SSL requires an appropriate design of the pre-text task to help the deep learning model capture essential features that assist ECG recovery, while no existing work has investigated only learning from radar signals without the aid of ECG ground truth. 

\section{Methodology}\label{sec:method}
\subsection{Overview of CFT-RFcardi Framework}
\begin{figure*}[tb] 
    \centering 
    \includegraphics[width=2\columnwidth]{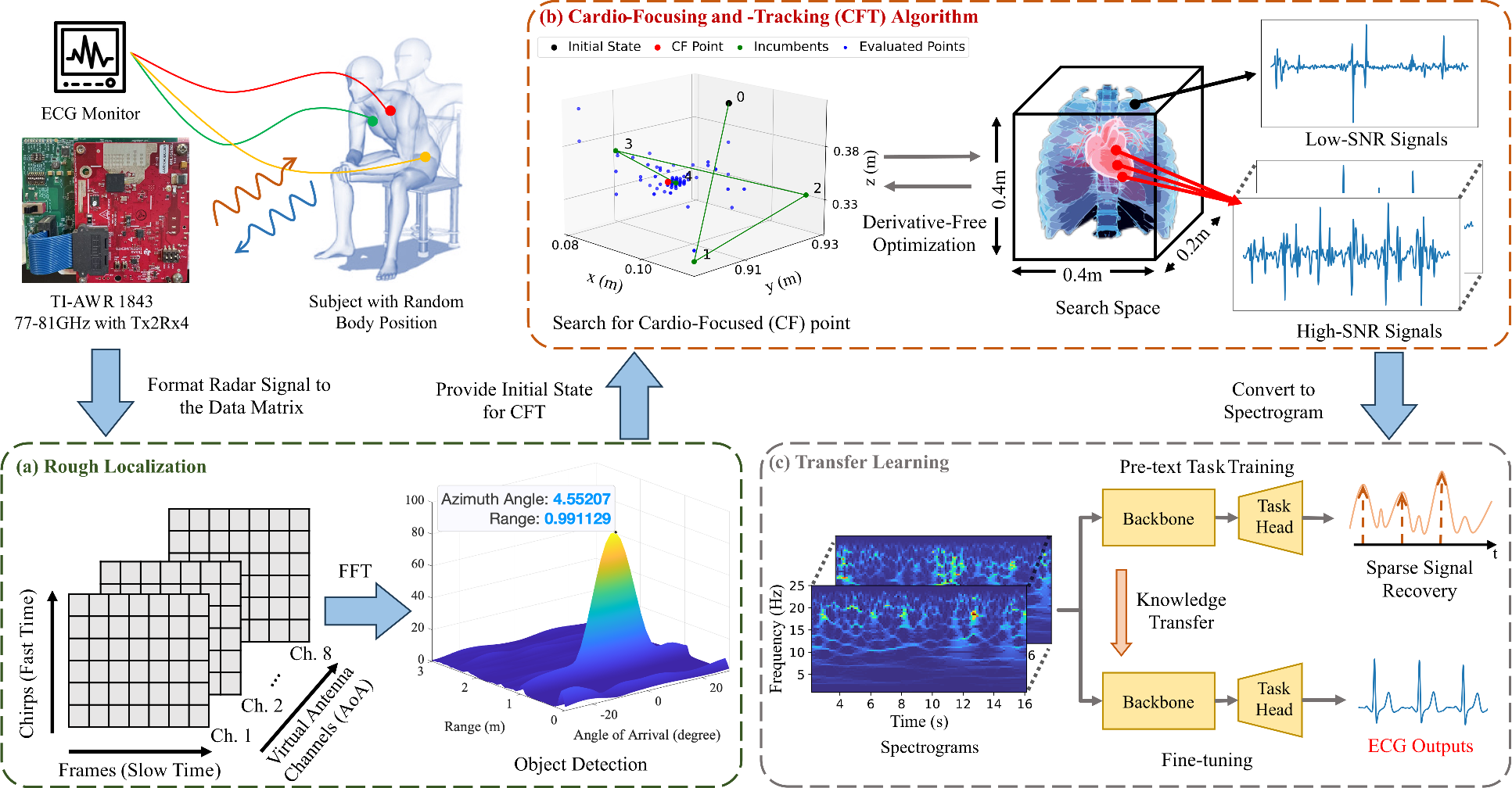}
    \caption{Overview of the CFT-RFcardi framework: (a) Rough localization of human body; (b) Use CFT to find CF point and extract high-SNR radar signals; (c) Transfer learning with pre-text task training and fine-tuning stages.}
    \label{fig:CFTRFcardi} 
\end{figure*}

The pipeline of the proposed CFT-RFcardi framework is shown in Figure~\ref{fig:CFTRFcardi} with three steps:
\begin{itemize}
\item The received radar signal will be converted into a standard format in terms of chirp, frame and virtual antenna channel to obtain the general location of the subject, as shown in Figure~\ref{fig:CFTRFcardi}(a).
\item The rough location acts as the initial state for the CFT algorithm, and the points within a constrained space will be evaluated to find the red CF point with the best SNR as shown in Figure~\ref{fig:CFTRFcardi}(b).
\item Signals extracted from the ten best points will be converted into spectrograms to pre-train the backbone with SSR as the pre-text task. Then, the same backbone will be used for the fine-tuning stage so that the latent representations learned in the pre-trained model can be seamlessly transferred for the ECG recovery task, as shown in Figure~\ref{fig:CFTRFcardi}(c).
\end{itemize}

In addition, TI-AWR 1843 radar operated at $77$GHz with $2$~Tx and $4$~Rx will be used for data collection, enabling $8$ virtual antenna channels for high-quality signal extraction. The detailed scenario descriptions and radar configurations will be provided in Section~\ref{sec:data_coll}. 

\subsection{Rough Localization}
The received radar signal is first formatted as a standard data matrix in terms of different chirps, frames and virtual channels to provide measurements of range, velocity and AoA, respectively. For the current research level of radar-based vital sign monitoring, the subjects are all quasi-static without velocity, and only the range-angle (RA) map will be calculated using fast Fourier transform (FFT) as shown in Figure~\ref{fig:CFTRFcardi}(a), with a detailed illustration of signal waveform and processing shown in Figure~\ref{fig:rouch_loc}.

\subsubsection{Range FFT}
According to (\ref{equ:raw_sig}), the signal propagation after transmitting introduces a constant phase shift $\phi_s$ in the received signal and is expressed as 
\begin{equation}\label{equ:range}
\phi_s = \frac{4\pi d_0}{\lambda}
\end{equation}
with $d_0$ representing the distance between radar and human body. Therefore, the distance $d_0$ can be extracted from each received signal along fast time using FFT as shown in Figure~\ref{fig:rouch_loc}(a), and the updated data matrix now reveals the range information, i.e., a static object denoted as blue along slow time axis.

\subsubsection{Angle FFT}
\begin{figure}[tb] 
    \centering 
    \includegraphics[width=1\columnwidth]{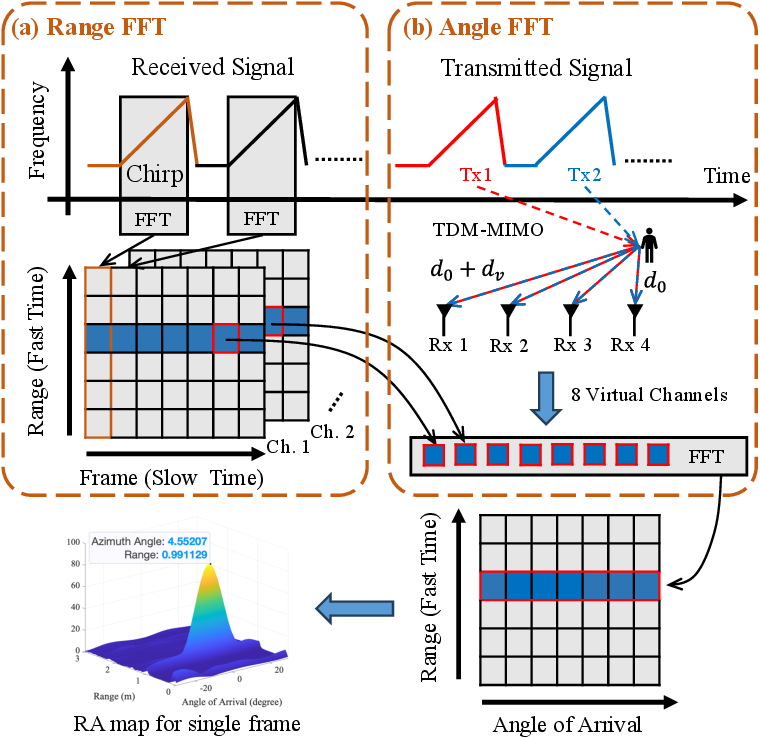}
    \caption{Procedures for obtaining RA map: (a) Range FFT for chirps along fast time; (b) Angle FFT along virtual channels.}
    \label{fig:rouch_loc} 
\end{figure}

The ability of AoA detection relies on the MIMO system using time division multiplexing (TDM-MIMO), with multiple Tx alternately transmitting chirp signals and the corresponding reflections can be distinguished during receiving as shown in Figure~\ref{fig:rouch_loc}(b). Due to the physical distance varies for different Tx/Rx combinations (i.e., Tx$2$Rx$4$ creates $8$ virtual channels), an extra propagation $\Delta \phi_v$ delay will be introduced as:
\begin{equation}\label{equ:angle}
\begin{aligned}
\Delta \phi_v &= \frac{4\pi d_v}{\lambda} \\
d_v &= l\ sin(\theta)
\end{aligned}
\end{equation}
where $d_v$ represents the extra propagation distance, $l$ means the distance between adjacent antenna channels and $\theta$ is the incident angle. Similar to range FFT, the phase differences across different channels can be used to extract AoA information for each range bin by performing FFT along the channel axis, as shown in red squares in Figure~\ref{fig:rouch_loc}(b).

After combining the FFT results for all chirps and channels, the final RA map for the current time sample (frame) can be obtained as shown in Figure~\ref{fig:rouch_loc}. The same procedure can be repeated along the slow time axis to get the rough human body location for all the time samples, but this study only requires the location obtained from the very first frame as the initial point $E_0$ for CFT algorithm.

\subsection{Cardio-focusing and -tracking (CFT) Algorithm}
The radar signal for any point can be extracted following (\ref{equ:raw_sig}) and (\ref{equ:phase}), and the search progress from $E_0$ to the best point $E_{b}$ (i.e., CF point with high SNR) requires: (a) an appropriate metric to assess whether the radar signal contains wanted cardiac features; (b) an optimization method that is applicable to the discontinuous objective space based on the assessed SNR values as costs. 

\subsubsection{Template Design for Assessing SNR}
\begin{figure}[tb]
  \centering
  \subfloat[]{\label{fig:target_org}\includegraphics[width=0.5\columnwidth]{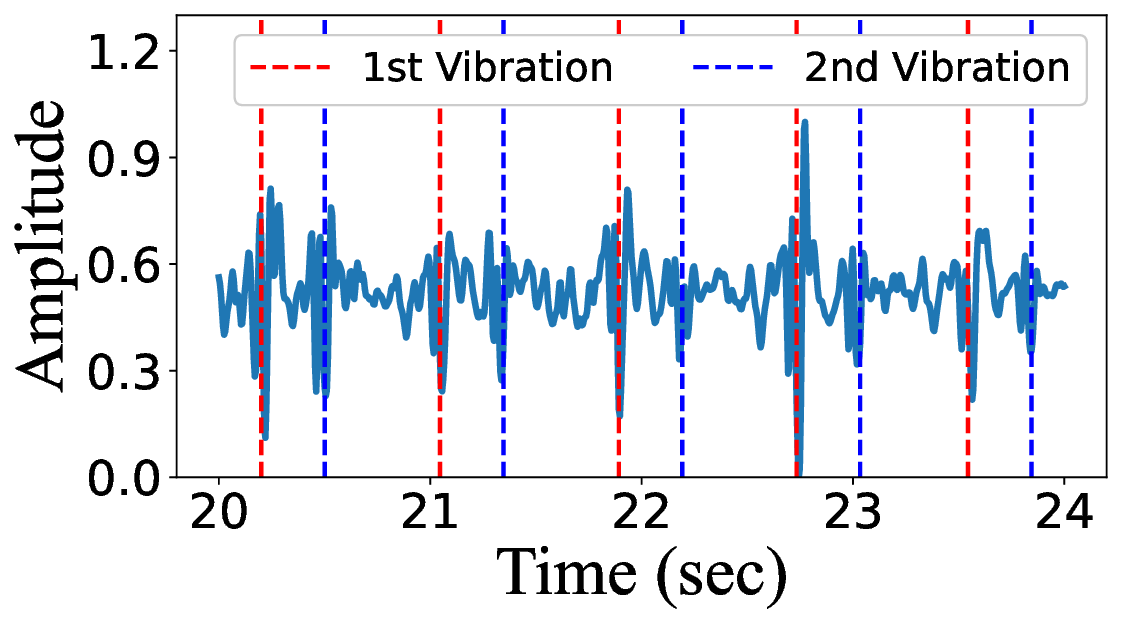}}
  \subfloat[]{\label{fig:target_sig}\includegraphics[width=0.5\columnwidth]{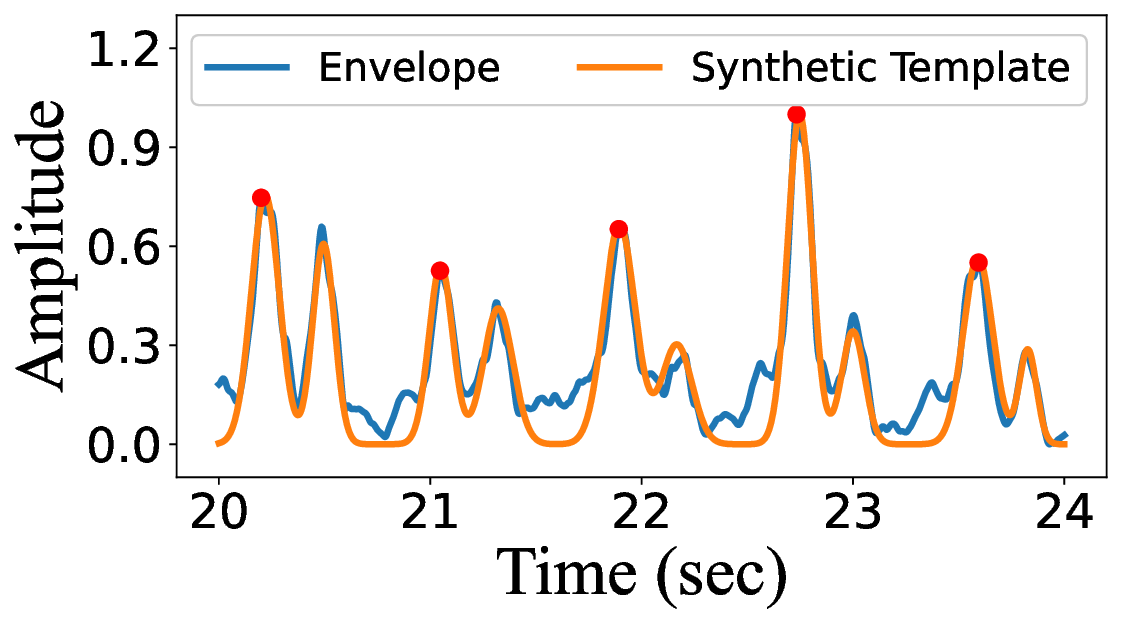}} \\
  \subfloat[]{\label{fig:target_org_bad}\includegraphics[width=0.5\columnwidth]{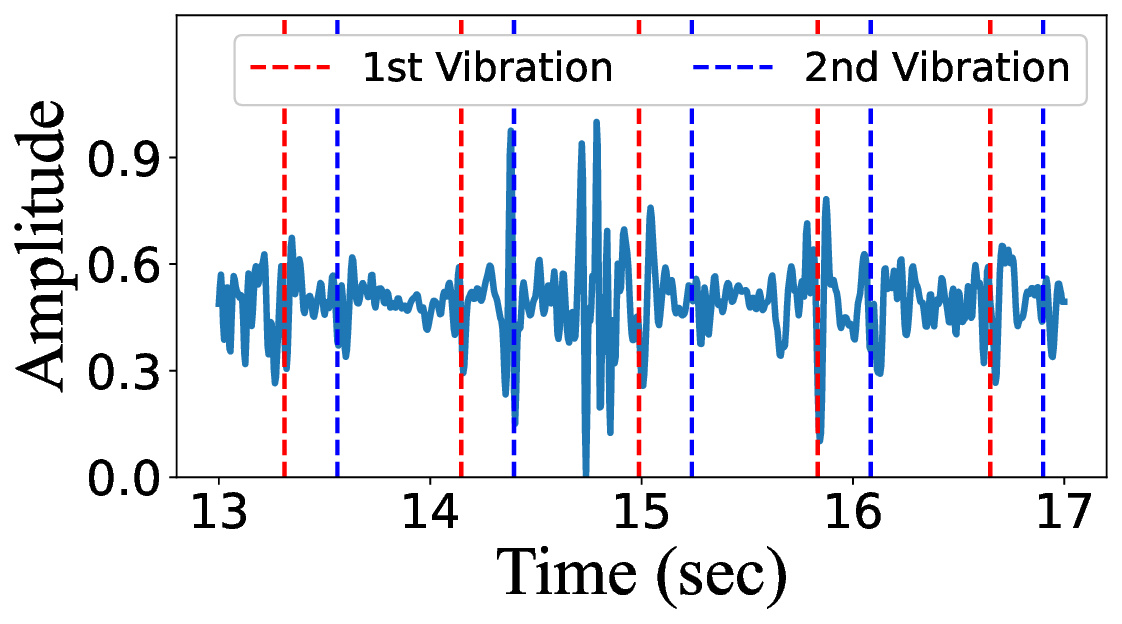}}
  \subfloat[]{\label{fig:target_sig_bad}\includegraphics[width=0.5\columnwidth]{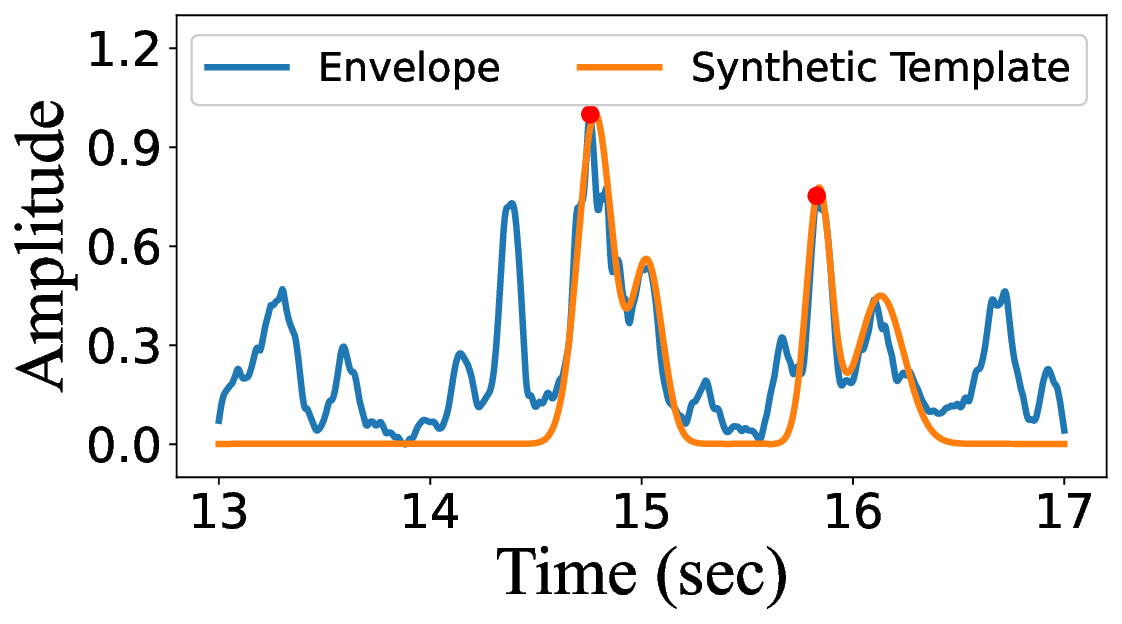}}
  \caption{Template for assessing SNR: (a) High-SNR radar signal; (b) Extracted signal envelope with the synthetic template; (c) (a) Low-SNR radar signal, (d) Extracted signal envelope with the synthetic template.}
  \label{fig:template}
\end{figure}

\textcolor{black}{An explicit SNR can be calculated with the known “clean” signal, while the “clean” signal for vital signs normally reveals two prominent vibrations corresponding to the ventricular contraction and relaxation~\cite{zhang2024radarODE}, as shown in Figure~\subref*{fig:target_org}.} However, considering the vibrations may have subtle differences due to different scenarios or radar configurations (e.g., noise figure and sampling frequency), a universal template $h_m$ is designed in this study to fit the envelope of the radar signal as: 
\begin{equation}
h_m (t) = a_1 \exp(-\frac{(t-b_1)^2}{2c_1^2}) + a_2 \exp(-\frac{(t-b_2)^2}{2c_2^2})
\end{equation}
with $a_1$, $a_2$ controlling the amplitudes of the peaks, $b_1$, $b_2$ determining the centers of the peaks and $c_1$, $c_2$ adjusting the width of the peaks. \textcolor{black}{In practice, $a_1$ and $b_1$ will be fixed based on the amplitude and position of the dominant peaks detected as the red points in Figure~\subref*{fig:target_sig}, and other parameters are left to be determined as a simple curve fitting problem as an optimization problem.} Finally, the mean square error (MSE) between the radar signal envelope and the synthetic template is reckoned to be an assessment of signal SNR as shown in Figure~\subref*{fig:target_sig}, because fewer components could fit the designed template for low-SNR radar signal without obvious cardiac features, as shown in Figure~\subref*{fig:target_org_bad} and~\subref*{fig:target_sig_bad}.

\subsubsection{Derivative-free Optimization (DFO)}
The MSE values obtained from template matching for the radar pieces extracted from point $E$ will be used as costs $\mathcal{F}(E)$ in searching CF point, but the traditional gradient-based optimization method is not applicable because there is no explicit cost function. Therefore, the CFT algorithm is developed in a derivative-free manner based on coordinate search (CS) algorithm~\cite{larson2019derivative}, to asymptotically approach the CF point. 

The definition of the DFO problem is formulated as:
\begin{equation}
E_b = \underset{E \in \mathbb{R}^n}{\arg \min }\left\{\mathcal{F}(E) : E\in \Omega \right\}
\end{equation}
with $\Omega$ representing a user-defined constrained $n$-dimensional search space near initial point $E_0$ as shown in Figure~\ref{fig:CFTRFcardi}(b), and the cost of points out of the constraint will be set as $\mathcal{F}(E\not\in \Omega)=\infty$. During each iteration $k$, many trial points $E_k$ within the constraint will be evaluated to find the incumbent points $E_i$ as the temporary best point for the next iteration. 

\begin{figure}[tb]
  \centering
  \subfloat[]{\label{fig:cft_1}\includegraphics[width=0.4\columnwidth]{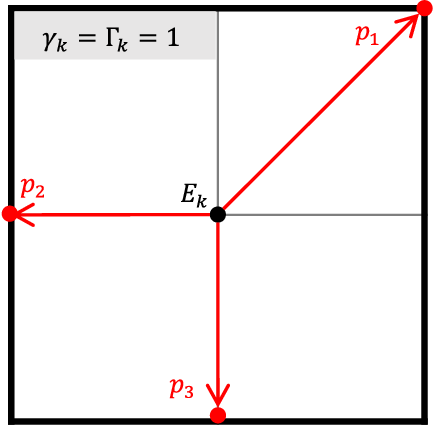}}\hspace{0.02\columnwidth}
  \subfloat[]{\label{fig:cft_2}\includegraphics[width=0.4\columnwidth]{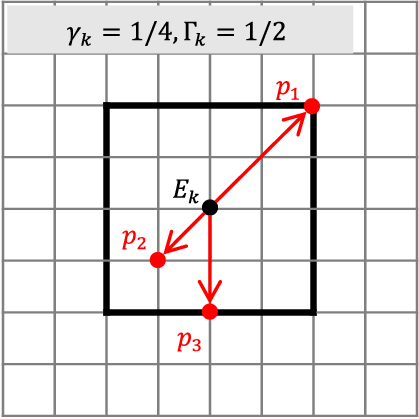}} \\
  \subfloat[]{\label{fig:cft_3}\includegraphics[width=0.7\columnwidth]{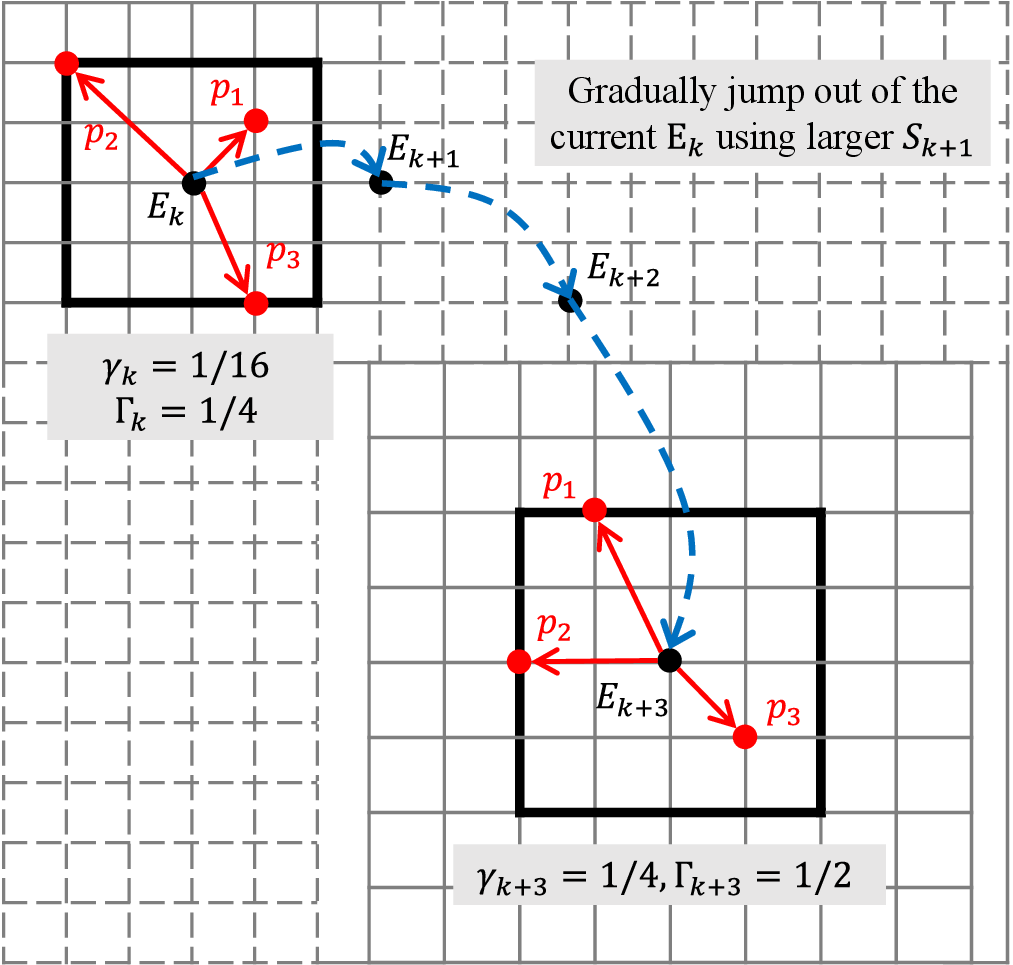}}
  \caption{Illustration of the CFT algorithm with bold line wrapping the search region $S_k$: (a) Equality between $\gamma$ and $\Gamma$ (same as in CS algorithm); (b) Large $\Gamma_k$ with refined $\gamma_k$, providing more potential points to be evaluated; (c) Jump out of the local minimum by adjusting $\Gamma_k$ and $\gamma_k$.}
  \label{fig:cft_plot}
\end{figure}

To perform a derivative-free search, the traditional CS algorithm starts from the initialization of grids $G_k$:
\begin{equation}\label{equ:grid}
G_k := \{E_k + \gamma_kD\} \subset \mathbb{R}^n
\end{equation}
where $\gamma_k>0$ is the grid size parameter and $D$ contains several vectors $p$ for possible searching directions, as shown in Figure~\subref*{fig:cft_1}. The local convergence of CS is ensured by dense search directions $D$ and a refined grid size $\gamma_k$ to find better $E_i$ compared with current $E_k$~\cite{larson2019derivative}. \textcolor{black}{As a basis for all CS-based DFO methods, $D$ can be formed by generating the orthogonal unit vectors based on the seed vector selected from the Halton sequence to ensure an evenly distributed search direction around the $E_k$.} However, the highly discontinuous objective space for radar-monitored vital signs may have numerous local minima that distract the optimization algorithm, i.e., the signal SNR of the adjacent points might be very different, as shown in Figure~\subref*{fig:radar_good} and~\subref*{fig:radar_bad}.

To jump out of the potential local minimum, CFT algorithm is proposed by introducing search region $S_k$ to restrict the possible search directions $p$, alleviating the difficulty of searching in numerous dense grid points and allowing the adjustment of search region and grid size iteratively to break the local minimum. The detailed procedures of CFT are shown in Algorithm~\ref{alg:CFT}, with an illustration of $\mathbb{R}^2$ space shown in Figure~\ref{fig:cft_plot}. 

In CFT, the grids $G_k$ is still expressed as in (\ref{equ:grid}) and the newly introduced search region $S_k$ is expressed as:
\begin{equation}
S_k := \{E\in G_k : \| E-E_k \|_{\infty} \leq \Gamma_k \}
\end{equation}
with $\Gamma_k$ as the size parameter for the search region. An intuitive interpretation of $S_k$ is the point set that contains grid points inside and on the boundary of the bold line controlled by $\Gamma_k$, as shown in Figure~\subref*{fig:cft_2}.

Based on the well-constructed grids $G_k$ and search region $S_k$, the remaining CFT algorithm is performed with searching and resizing stages:

\textbf{Searching: } The searching stage simply asks for the evaluation of $\mathcal{F}(E)$ on a subset of grids $G_k$ based on any sampling algorithm (e.g., Latin hypercube sampling~\cite{larson2019derivative}), as indicated in line~\ref{line:s1} in Algorithm~\ref{alg:CFT}.

\textbf{Resizing: } The resizing stage depends on the result of searching stage: 
\begin{itemize}
  \item If a new incumbent point $E_i$ is found with better SNR, the search region will be doubled as $\Gamma_{k+1}=2\Gamma_k$ (line~\ref{line:r1}), and the grid size will be empirically set as $\gamma_{k+1}=\min(\Gamma_k, {\Gamma_k}^2)$ (line~\ref{line:r4}), enabling the search in a broader space in the next iteration.
  \item If there is no better point than the current $E_k$ on the current grids $G_k$, another searching stage will be performed only within the search region $S_k$ (line~\ref{line:s2}). Then, If a better point $E_i$ is found, $\Gamma_{k+1}$ and $\gamma_{k+1}$ are obtained as above (line~\ref{line:r2}), otherwise, the search region will be halved as $\Gamma_{k+1}=\Gamma_k/2$ (line~\ref{line:r3}) for a finer search with $\gamma_{k+1}=\min(\Gamma_k, {\Gamma_k}^2)$ (line~\ref{line:r4}).
\end{itemize}

The searching step enables the finding of better points $E_i$ in a broad space, and the resizing step either refines the grid if the current $\gamma_{k}$ is not enough or enlarges the search space when stalling at the local minimum, as shown in Figure~\subref*{fig:cft_3}. Finally, the CFT algorithm will be terminated after achieving a desired SNR$_d$ or iteration limit $k_{max}$.

\begin{algorithm}[tb]
\caption{CFT Algorithm}\label{alg:CFT}
\begin{algorithmic}[1]
    \State \textbf{Input:} $E_0$, SNR$_d$, $k_{max}$
    \State \textbf{Output:} $E_b$, SNR$_b$
    \Statex \textsc{Objective}:
    \State $E_b = \underset{E \in \mathbb{R}^n}{\arg \min }\left\{\mathcal{F}(E) : E\in \Omega \right\}$
    \State Initialize $k=0$, $\Gamma_k = \gamma_k = 1$, $SNR_b=\mathcal{F}(E_k)$
    \While {$SNR_b<SNR_d$ and $k<k_{max}$}
    \If{$\mathcal{F}(E)<SNR_b$ for some $E\in G_k$} \label{line:s1}
      \State $E_{k+1}\leftarrow E$, $SNR_b \leftarrow \mathcal{F}(E)$ 
      \State $\Gamma_{k+1}\leftarrow 2\Gamma_k$ \label{line:r1}
    \ElsIf{$\mathcal{F}(E)<SNR_b$ for some $E\in S_k$} \label{line:s2}
      \State $E_{k+1}\leftarrow E$, $SNR_b \leftarrow \mathcal{F}(E)$
      \State $\Gamma_{k+1}\leftarrow 2\Gamma_k$ \label{line:r2}
    \Else
      \State $E_{k+1} \leftarrow E_{k}$
      \State $\Gamma_{k+1}\leftarrow \Gamma_k/2$ \label{line:r3}
    \EndIf
    \State $\gamma_{k+1}\leftarrow \min(\Gamma_k, {\Gamma_k}^2)$ \label{line:r4}
    \State $k \leftarrow k+1$, $E_b \leftarrow E_{k+1}$
    \EndWhile
\end{algorithmic} 
\end{algorithm} 

The visualization of the CFT algorithm in Figure~\ref{fig:CFTRFcardi}(b) shows that the initial few iterations search in a large space, and the algorithm could jump out of the green local minima to find the red CF point within the fine blue grid points. In addition, the tracking of the CF points along time can be naturally realized by repeating Algorithm~\ref{alg:CFT} with previous $E_b$ as the new $E_0$, and the SNR evaluated on the previous point might have already achieved SNR$_d$ due to the quasi-static human body, saving a huge amount of time for calculating useless channel information for filtering or clustering~\cite{li2024radarnet,chen2022contactless,liu2024diversity}. 

\subsection{Transfer Learning for ECG Recovery}
\subsubsection{Deep Learning Model Design}
The radar signal extracted from the $10$ best points from CFT algorithm will be converted to spectrograms as the input for deep learning model according to our previous work~\cite{zhang2024radarODE-MTL}, providing extra frequency-domain information to assist model training. The deep learning model adopts the popular backbone-decoder structure as designed in~\cite{zhang2024radarODE-MTL}:
\begin{itemize}
  \item The backbone leverages ResNet~\cite{he2020resnet} framework with deformable 2D convolution layer~\cite{dai2017deformable} to efficiently extract cardiac features from image-like spectrogram inputs.
  \item The decoder is based on 1D convolutional neural network (CNN) to generate corresponding signals either for the pre-text task or ECG signal recovery. 
\end{itemize}

\subsubsection{Pre-text Task Training and Fine-tuning}
\color{black}
The pre-text task used for SSL should reveal certain inherent features in the radar-monitored vital sign and can be transferred to help the ECG recovery. According to our previous work~\cite{zhang2024radarODE-MTL}, two major features of a good ECG recovery are accurate morphological features and accurate peak locations. The learning of morphological features fully relies on the non-linear mapping ability of the DL model by training with massive ECG data and is impossible to be pre-trained using radar signal only. However, the peaks in radar (heartbeats) and ECG signals (R peaks) are synchronized and have been extracted when assessing the radar signal SNR, as shown in Figure~\subref*{fig:target_sig}.

It is natural to leverage the task for heartbeat detection as the pre-text task for SSL. Two categories of methods used for traditional heart rate estimation are based on periodicity and sparsity~\cite{zhang2023overview}. In this work, the duration of each segment is $4$ sec and may not reveal strong periodicity. Therefore, SSR will be used as the pre-text task in RFcardi and is defined as:
\begin{equation}
h=\Phi x+n
\end{equation}
\color{black}
where $h$ is the high-SNR radar signal, $\Phi$ means the observation matrix, $x$ is the sparse representation for heartbeats and $n$ represents the residual noise. The traditional SSR task can be seamlessly converted to a system identification problem by viewing $\Phi$ as a multi-channel adaptive filter, and the estimation of the filter coefficient is the same as training a CNN-based neural network (i.e., training CNN kernels)~\cite{ha2020contactless}.

In this case, the SSR task is realized by using the aforementioned CNN-based backbone-decoder structure with the loss function:
\begin{equation}\label{equ:sparse}
\mathcal{L} = \|x - \tilde{x} \|_2 + \underbrace{\lambda_s\frac{\|x\|_1 / \| x\|_2-1}{\sqrt{m}-1}}_{\text{sparse penalty}}
\end{equation}
where $m$ is the length of the signal, $x$ is the output from deep learning model, $\tilde{x}$ is the sparse ground truth with values for the radar peaks maintained (1st vibration in Figure~\subref*{fig:target_org}) while other values are set to $0$. The sparse penalty has a range of $[0,\lambda_s)$ with a smaller value indicating better sparsity~\cite{hoyer2004non}.

After pre-training based on SSR, the parameters of backbone will be retained with a new decoder connected (same structure as for pre-text task training), and a few radar-ECG pairs are used for fine-tuning the pre-trained RFcardi model using MSE as the loss function.

\section{Details of Experiment and Dataset}\label{sec:exp}
\subsection{Experimental Details}\label{sec:data_coll}
\color{black}
\subsubsection{Dataset Collection and Preparation}
\begin{figure}[tb]
  \centering
  \includegraphics[width=1\columnwidth]{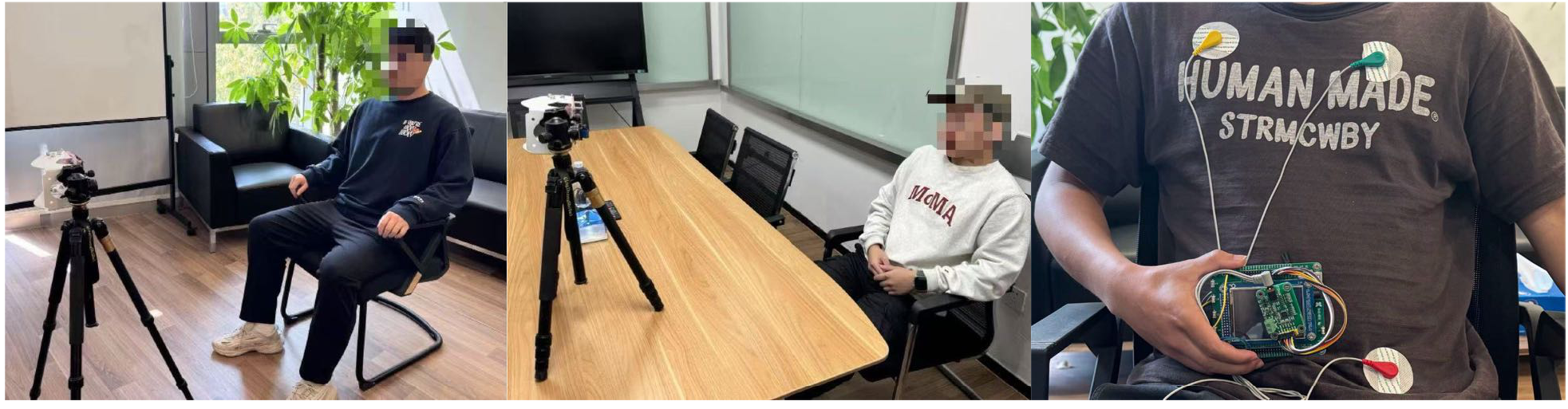}
  \caption{\textcolor{black}{Indoor scenarios for radar signal collection, with the synchronous ECG ground truth collected.}}
  \label{fig:data_col}
\end{figure}

The dataset contains a total of $80$-minute synchronous radar-ECG pairs collected for $5$ healthy subjects ($3$ men, $2$ women) in $2$ indoor scenarios as shown in Figure~\ref{fig:data_col}. The subjects are asked to sit casually and are allowed to change postures during data collection, and each data trial lasts for $1$ minute. The upper body travel distances vary from $1$ to $9$m during $1-$minute trials based on the statistics analyzed after the experiment. The distance between radar and human body varies from $0.5$ to $1.2$m, and a longer distance causes a decrease in signal SNR, with a smaller portion of the space points containing useful cardiac features.

TI-AWR 1843 radar with $2$ Tx and $4$ Rx is used for data collection with $8$ virtual antenna channels created~\cite{AWR1843}, and the radar configurations are listed in Table~\ref{tab:data_param} with the names provided in TI mmWave-Studio interface. Meanwhile, the ECG ground truth is collected using TI ADS1292, with AC coupling and an integrated right-leg drive (RLD) amplifier to remove potential baseline drift or power-line noise. The related ECG processing (e.g., smoothing and peak finding) is realized by NeuroKit2 Python package~\cite{makowski2021neurokit2}. All the signals are sampled at $200$Hz, and only a band-pass filter from $0.5$ to $50$Hz and a differentiator are used for removing respiration noise because the radar signal extracted from CF points already has high SNR. Lastly, the CFT algorithm is run on the Apple M2 Pro CPU with $32$GB RAM.
\color{black}

\begin{table}[tb]
\centering
\color{black}
\caption{Parameters for data collection interface}
    \begin{tabular}{lc?lc}
    \toprule
    \textbf{Parameter} & \textbf{Value} & \textbf{Parameter} & \textbf{Value} \\
    \toprule
    Start Frequency & $77$GHz & Frequency Slope & $65$MHz/$\mu$s \\ 
    Idle Time & $10\mu$s    & Tx Start Time & $1\mu$s \\
    ADC Start Time & $6\mu$s  & ADC Samples & $256$ \\ 
    Sample Rate & $5000$kbps  & Ramp End Time & $60\mu$s \\ 
    Start/End Chirp Tx & $0/2$  & No. of Chirp Loops & $2$  \\ 
    No. of Frames & $12000$  & Frame Periodicity & $5$ms  \\ 
    Tx Gain ($G_T$) & $10$ dBi  & Rx Gain ($G_R$) & $30$ dBi  \\
    Tx power ($P_{\mathrm{T}}$)& $12$ dBm & Noise Figure (NF)     & $15$ dB   \\ 
    Wavelength ($\lambda$)  & $3.9$ mm               & Bandwidth (B)         & $3.8$ GHz \\
    \bottomrule
    \end{tabular}
\label{tab:data_param}
\end{table}%

\color{black}
\subsection{Implementation Details}
\subsubsection{Parameters for CFT Algorithm}
The constraint $\Omega$ for the CF point search is centered at the initial state $E_0$ with a range of $0.4\times 0.2 \times 0.4$m as illustrated in Figure~\ref{fig:CFTRFcardi}(b). In addition, the initial grid and search region size should be adjusted to fit the real-life physical unit as $\Gamma_k= \gamma_k = 0.1$m, and the size will be limited as $\Gamma_k\geq\gamma_k\geq 0.001$m to prevent an exhaustive search within a meaningless small space. Then, SNR$_d$ is empirically set to $0.01$ for the desired MSE between the normalized synthetic template and the signal envelope, ensuring that the extracted signal contains enough cardiac features for ECG recovery. At last, $k_{max}$ is empirically set to $100$ to balance the quality and cost of the space search.

\subsubsection{Deep Learning Model Training}
The deep learning model adopts the same backbone, ECG decoder and hyperparameters as in our previous open-sourced work~\cite{zhang2024radarODE-MTL} coded in PyTorch and trained on NVIDIA RTX 4090 (24GB), and the total training epoch is set to $100$ with batch size $8$. All the experiments are repeated 5 times, and the mean values are taken to ensure the results are statistically reliable, but the confidence interval is not provided because all the improvements show statistical significance.

The dataset is split based on a 5-fold cross-validation training strategy, with the trials from 1 fixed subject for testing and the other 4 subjects alternately selected for training or validation, ensuring to make use of all the possible trials while not involving the testing data in the training phase. During each training iteration, the 4-sec input radar segment is split from each 1-minute trial with a step length of 1~sec, generating a total of 4560 samples for training, evaluation and testing.
\begin{table*}[tb]
\centering
\color{black}
\caption{Comparison table for different methods}
    \begin{tabular}{c|cccc?c|ccc}
    \toprule
    \makecell[c]{Methods \\ (Improve SNR)} & Feature & \makecell[c]{Vital \\ Sign} & \makecell[c]{Real-\\time} & \makecell[c]{Allow \\ posture change} & \makecell[c]{Methods \\ (SSL)} & Input & Output & Loss\\
    \midrule
    De-ViMo$^*$~\cite{liu2024diversity} & Accumulation & HR & \ding{52} & \ding{55} & MaeFE~\cite{zhang2022maefe} & Masked radar signal & Original radar signal  & MSE \\
     MMECG~\cite{chen2022contactless} & Clustering & ECG & \ding{55} & \ding{55} & CL~\cite{canellas2025self} & Pos. and neg. samples &  Latent vectors &  NT-Xent~\cite{canellas2025self} \\
     CFT  & DFO & ECG & \ding{55} & \ding{52} & RFcardi & Radar signal & Sparse heartbeats & MSE \\
    \bottomrule
    \multicolumn{6}{l}{$^*$For the original De-ViMo with only HR monitoring ability.}
    \end{tabular}%
\label{tab:comp_table}
\end{table*}%
\subsection{Methods for Comparison}
The comparison is performed with the representative methods based on accumulation and clustering to extract high-SNR radar signals:
\begin{itemize}
  \item De-ViMo~\cite{liu2024diversity} is proposed for real-time heart rate (HR) monitoring and is based on the accumulation of signals from various dimensions (i.e., $2$ chirps, $8$ virtual antennas, $10$ spatial points) to enhance cardiac features while mitigating noise. However, original De-ViMo only generates a sinusoidal wave, which cannot be used for ECG recovery. Therefore, an extra signal extraction stage is needed, but the real-time monitoring will be unachievable.
  \item MMECG~\cite{chen2022contactless} requires the calculation of numerous points in 3D space and applies clustering algorithm to improve SNR. Then, a pattern-matching process is performed to learn the common pattern from the clustered result and select the best radar signal(s). In this study, the number of 3-D points used for clustering is $6\times 6\times 6 = 216$.
\end{itemize}

In addition, to the best of our knowledge, this is the first paper that leverages SSL with an appropriate pre-text task for pre-training of radar-based ECG recovery. Therefore, two widely applicable SSL methods for biological signals are used for comparison:
\begin{itemize}
  \item MaeFE~\cite{zhang2022maefe} is designed based on the famous masked autoencoder (MAE), with masked reconstruction of signal patches as the pre-text task. To fit our task, each 4-sec input radar signal will be masked using faint Gaussian noises ($3$~dB) with a length of $1$~sec.
  \item The method proposed in~\cite{canellas2025self} is based on contrastive learning (CL) by constructing positive and negative pairs and maximizing similarity between the latent vectors generated for positive pairs while minimizing similarity for negatives. To fit our task, the positive pairs are built by the records taken from two different channels in the same 4‑sec window, while the negative pairs are formed by pairing a window with another window more than 20 seconds apart in time, ensuring a different physiological state driven by changing activity.
\end{itemize}
The core features of all the methods used in this paper have been summarized in Table~\ref{tab:comp_table} for clarity.

\subsection{Evaluation Metrics}
The performance of the proposed CFT-RFcardi framework will be separately evaluated in terms of the quality (SNR) of the collected radar signal and the recovered ECG signal using the following metrics:
\begin{itemize}
  \item For both radar collection and recovered ECG signals, the peak accuracy is crucial because it links to many important cardiac features (e.g., heart rate variability) for clinical diagnosis. Thus, absolute \textbf{peak error} between ECG R peaks and the dominant peaks for the first vibrations (i.e., heartbeats) in the radar or recovered ECG signal is used to measure the peak accuracy.
  \item Another metric related to peaks is \textbf{missed detection rate (MDR)} to count the cardiac cycles with no peak detected or with the absolute peak error larger than $150$ms~\cite{chen2022contactless}, providing a quantified evaluation of the corrupted radar signal collection or ECG recoveries due to the noise.
  \item For ECG recovery, the morphological accuracy is measured using \textbf{MSE} and \textbf{Pearson-correlation coefficient (PCC)} as calculated in (\ref{equ:mse_pcc}), with MSE sensitive to the peak deviation and PCC focusing on the similarity between the ECG patterns.

\begin{equation}\label{equ:mse_pcc}
\begin{aligned}
\mathrm{MSE}&=\frac{1}{N} \sum_{i=1}^N\left(x_i-x_i^{\prime}\right)^2 \\
\operatorname{PCC}&=\frac{\sum_{i=1}^N\left(x_i-\bar{x}\right)\left(x_i^{\prime}-\bar{x}^{\prime}\right)}{\sqrt{\sum_{i=1}^N\left(x_i-\bar{x}\right)^2} \sqrt{\sum_{i=1}^N\left(x_i^{\prime}-\bar{x}^{\prime}\right)^2}}
\end{aligned}
\end{equation}
where the predicted and ground truth ECG pieces are $x$, $x^{\prime}$, with length $N$ and mean values $\bar{x}$, $\bar{x}^{\prime}$.
\end{itemize}
\color{black}

\begin{figure*}[tbp]
  \centering
  \begin{minipage}{0.1\linewidth}\centering
\rotatebox[origin=center]{0}{\textbf{Rough Loc.}}
\rotatebox[origin=center]{0}{$\approx$}\\
\rotatebox[origin=center]{0}{\textbf{CF Point}}
\end{minipage}\begin{minipage}{0.9\linewidth}\centering
  \subfloat[]{\label{fig:bf_good}\includegraphics[width=0.3\columnwidth]{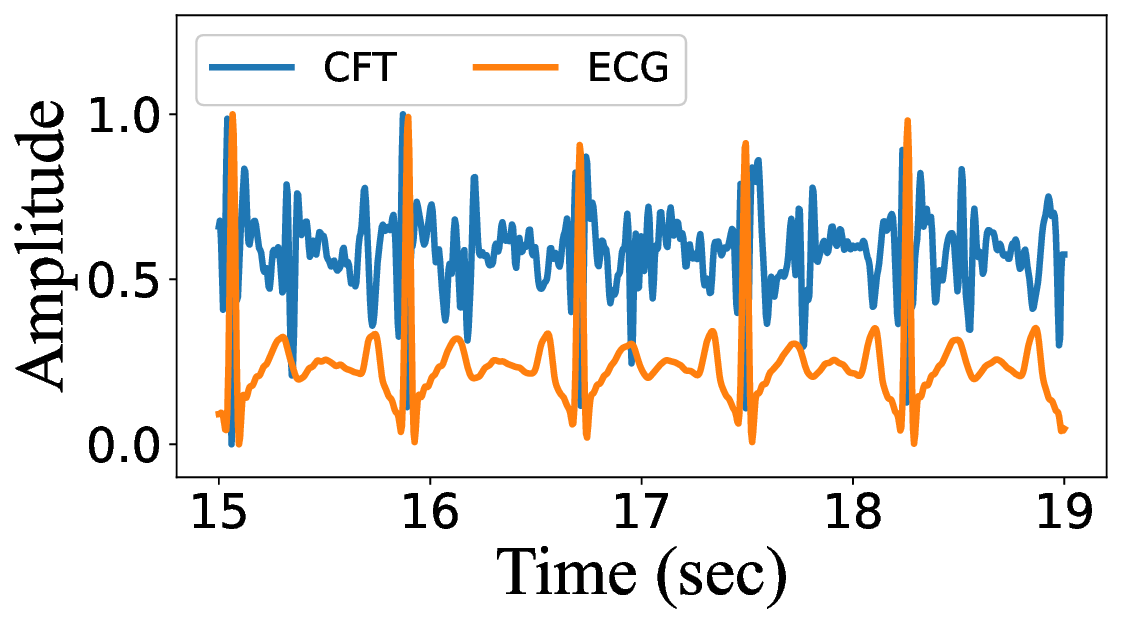}}
  \subfloat[]{\label{fig:MMECG_good}\includegraphics[width=0.3\columnwidth]{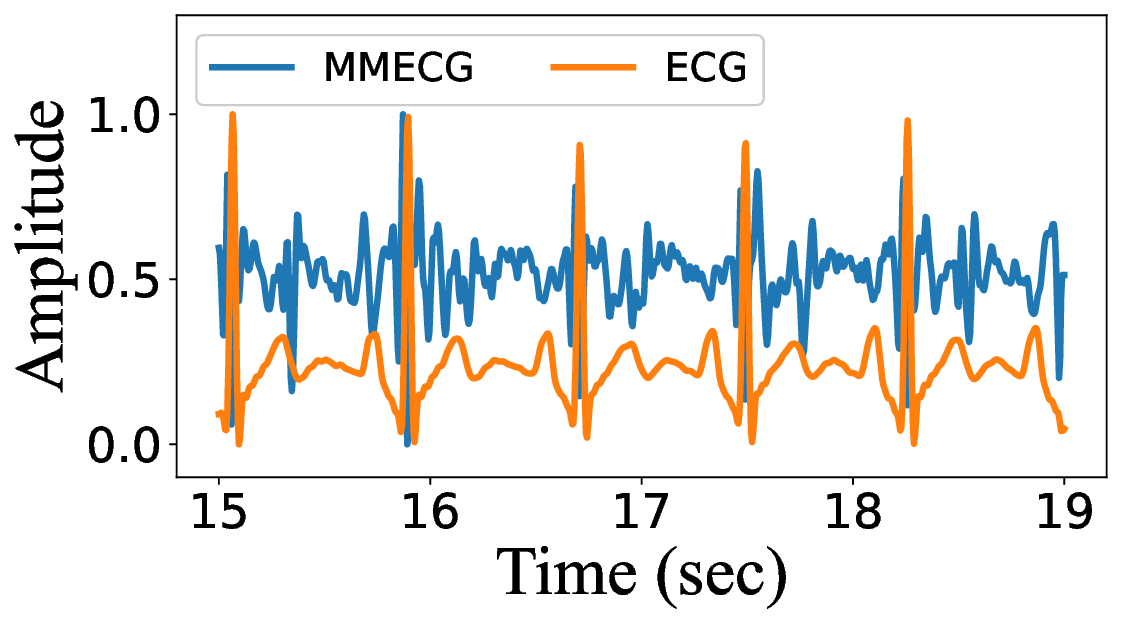}}
  \subfloat[]{\label{fig:vimo_good}\includegraphics[width=0.3\columnwidth]{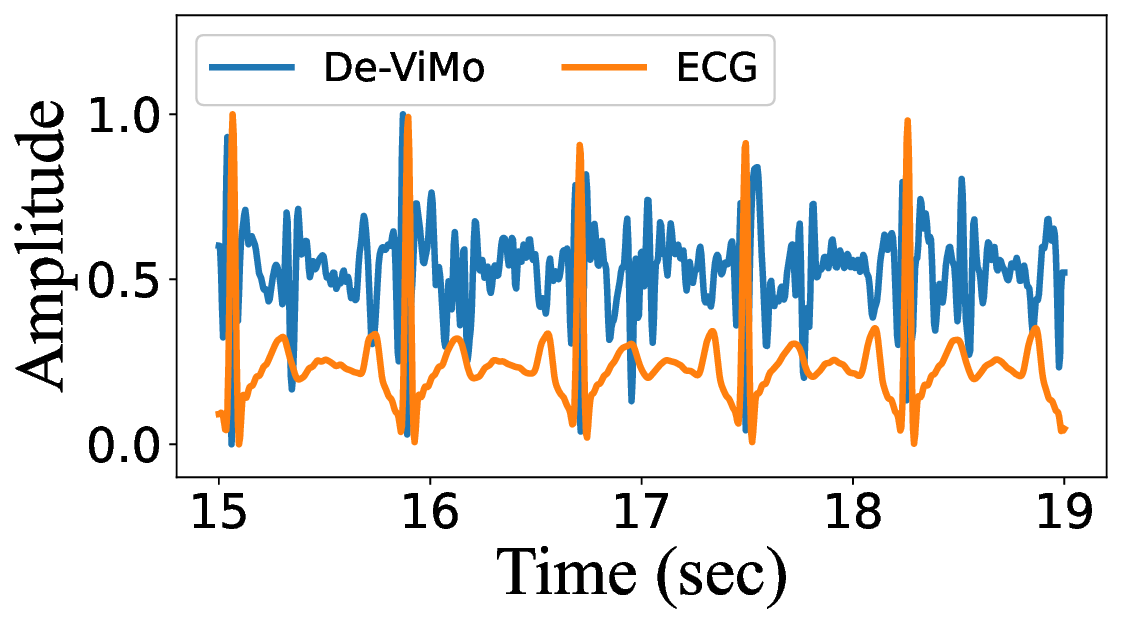}}\\
\end{minipage}
\begin{minipage}{0.1\linewidth}\centering
\rotatebox[origin=center]{0}{\textbf{Rough Loc.}}
\rotatebox[origin=center]{0}{$\neq$}\\
\rotatebox[origin=center]{0}{\textbf{CF Point}}
\end{minipage}\begin{minipage}{0.9\linewidth}\centering
  \subfloat[]{\label{fig:bf_bad}\includegraphics[width=0.3\columnwidth]{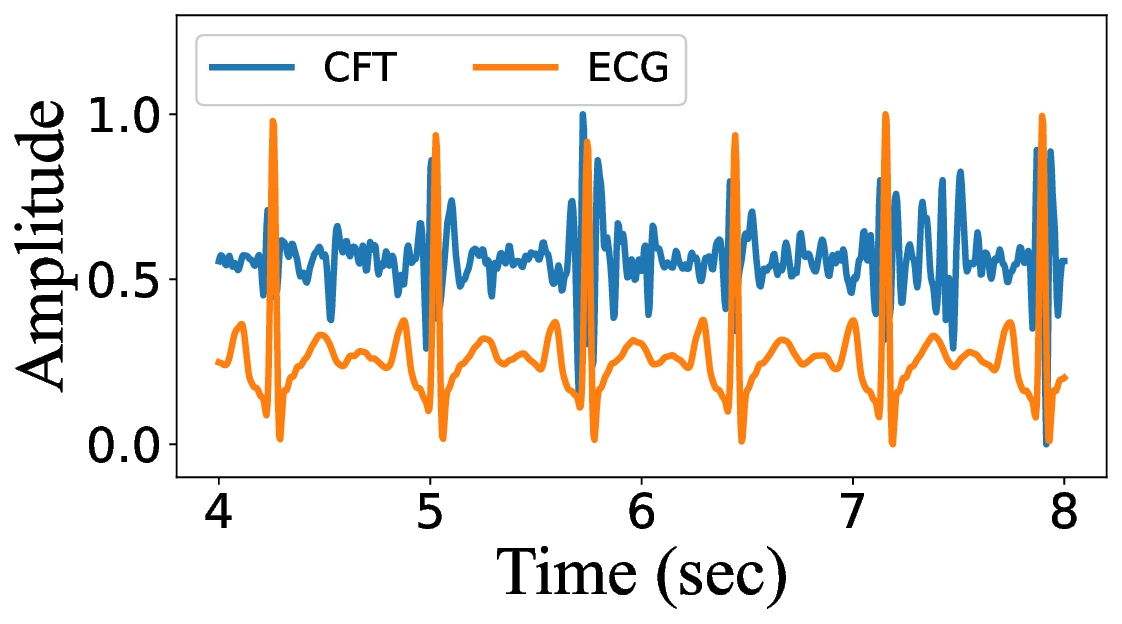}}
  \subfloat[]{\label{fig:MMECG_bad}\includegraphics[width=0.3\columnwidth]{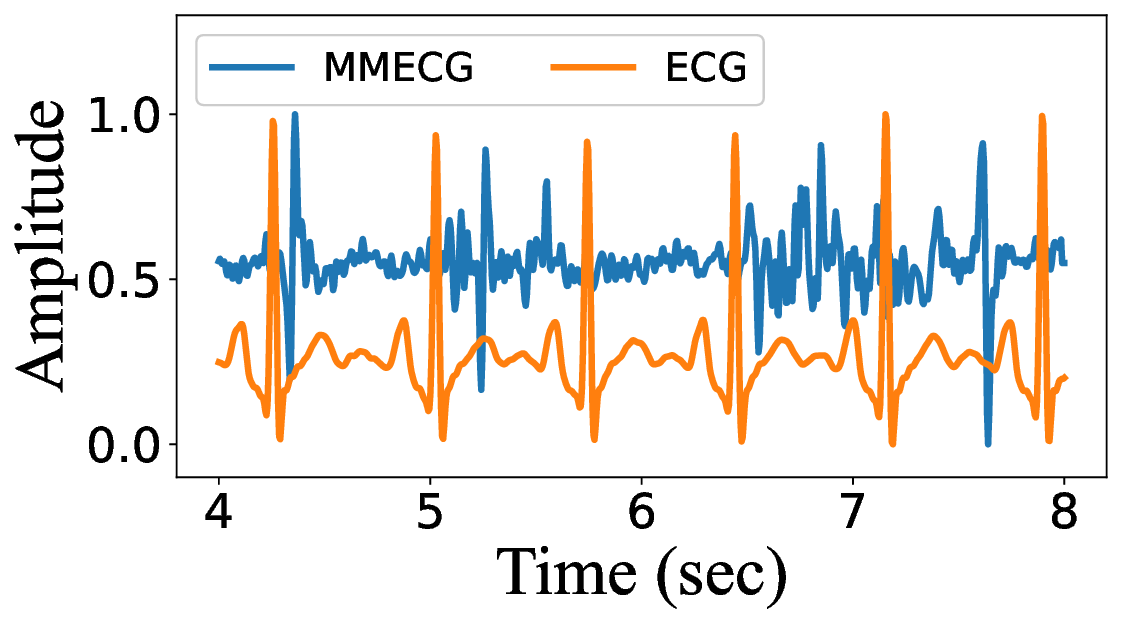}}
  \subfloat[]{\label{fig:vimo_bad}\includegraphics[width=0.3\columnwidth]{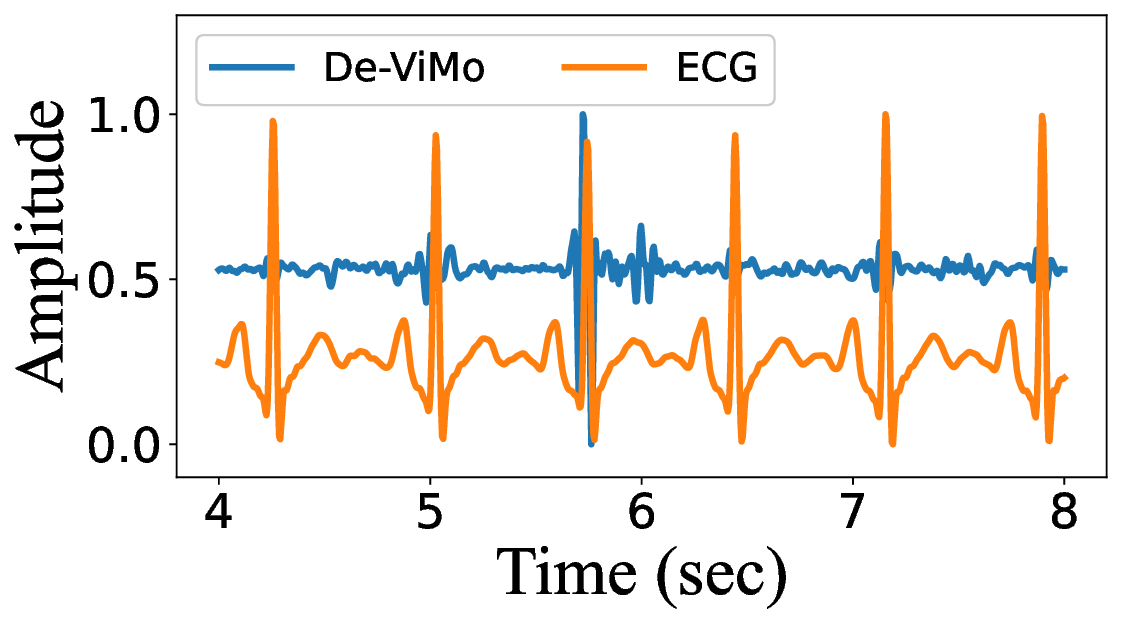}}
\end{minipage}
  \caption{Visualization of the extracted radar signal for all methods: (a) - (c) If CF point is around rough body location; (d) - (f) If CF point is far from rough body location.}
  \label{fig:compare_overall}
\end{figure*}

\section{Experimental Results and Evaluations}\label{sec:result}
\subsection{Performance of CFT Algorithm}
\subsubsection{Effectiveness of CFT Algorithm}
The examples of the extracted radar signal for different methods are shown in Figure~\ref{fig:compare_overall}, illustrating that precise cardiac localization has a huge effect on the signal quality. For example, if the rough body location is around CF point, all three methods can obtain high-SNR signals with clear first and second vibrations using either space search (CFT, Figure~\subref*{fig:bf_good}), clustering (MMECG, Figure~\subref*{fig:MMECG_good}) or accumulation (De-ViMo, Figure~\subref*{fig:vimo_good}). 

In contrast, only a few range bins will contain useful cardiac features if the rough body location is far from CF points, especially when increasing the monitoring range. Therefore, the signal accumulation may enhance the noises as shown in Figure~\subref*{fig:vimo_bad} while the signal clustering may also encounter a failure due to the lack of homogeneous cardiac signals as shown in Figure~\subref*{fig:MMECG_bad}. However, The proposed CFT can precisely locate the CF point with good SNR subject to the designed signal template and DFO searching strategy, and the extracted radar signal still shows clear peaks as shown in Figure~\subref*{fig:bf_bad}.

During the data collection of this study, the subjects are allowed to change postures to alleviate discomfort, with a resultant CF point deviation of several decimeters, while the rough location provided by FMCW signal processing is still unchanged. Therefore, the proposed CFT algorithm is essential because the posture change is inevitable, and a thorough evaluation in terms of different monitoring ranges will be performed in the next part.

\color{black}
\subsubsection{Impact of Monitoring Range}
\begin{figure}[tb]
  \centering
  \subfloat[]{\label{fig:pk_err_point}\includegraphics[width=0.5\columnwidth]{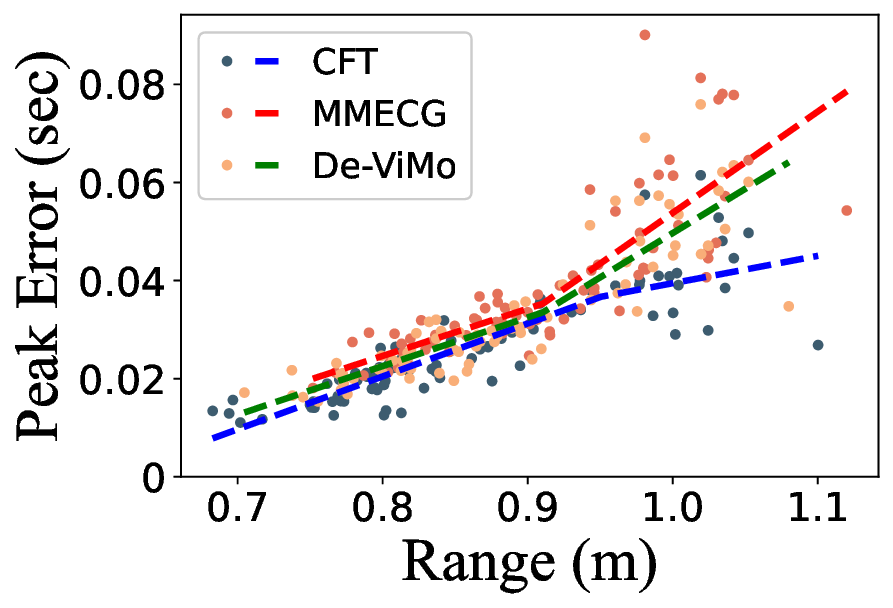}}
  \subfloat[]{\label{fig:pk_mdr_point}\includegraphics[width=0.48\columnwidth]{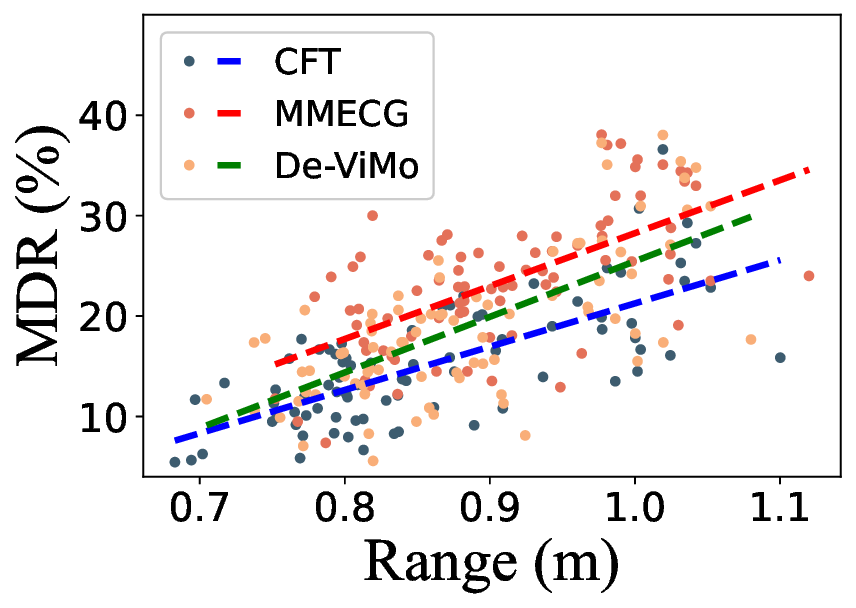}} \\
  \subfloat[]{\label{fig:pk_err_point_dis}\includegraphics[width=0.5\columnwidth]{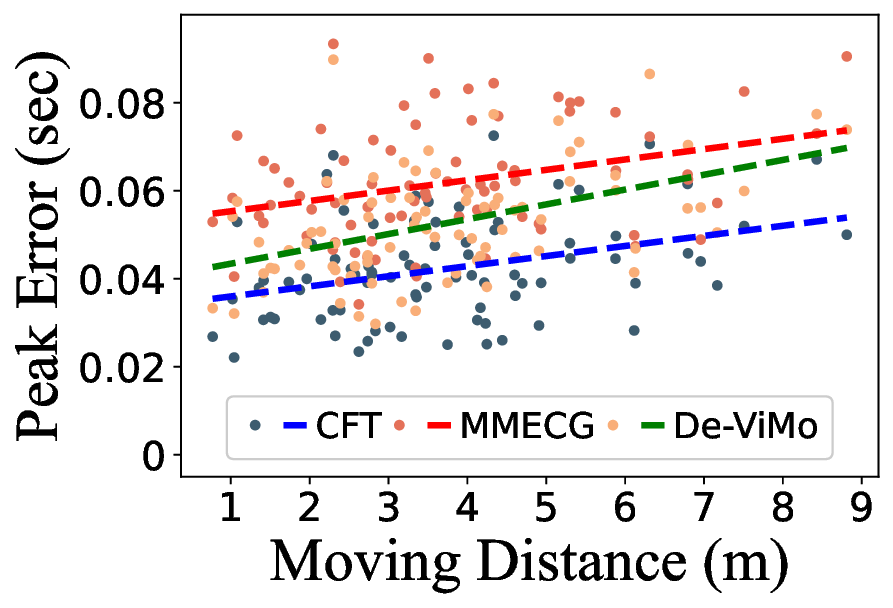}}
  \subfloat[]{\label{fig:pk_mdr_point_dis}\includegraphics[width=0.48\columnwidth]{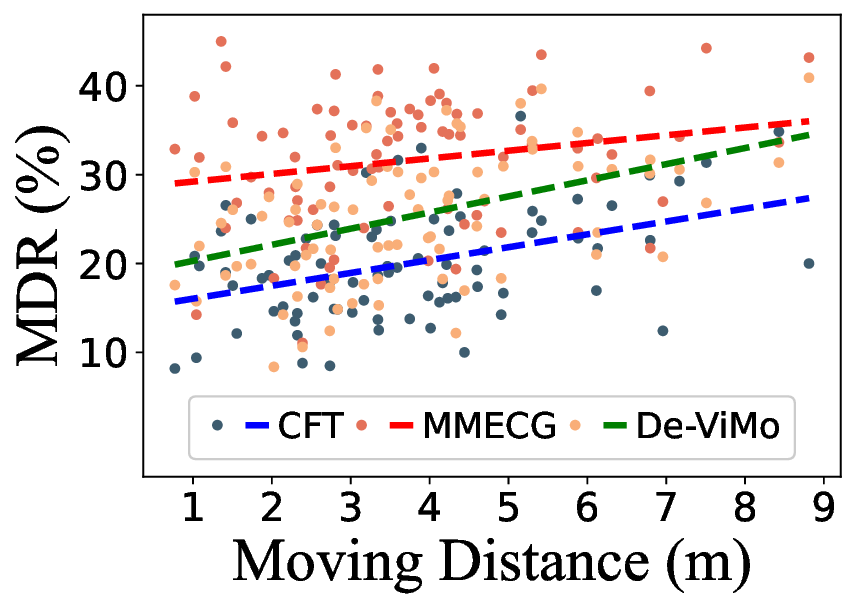}} \\
  \subfloat[]{\label{fig:pk_err_cdf}\includegraphics[width=0.5\columnwidth]{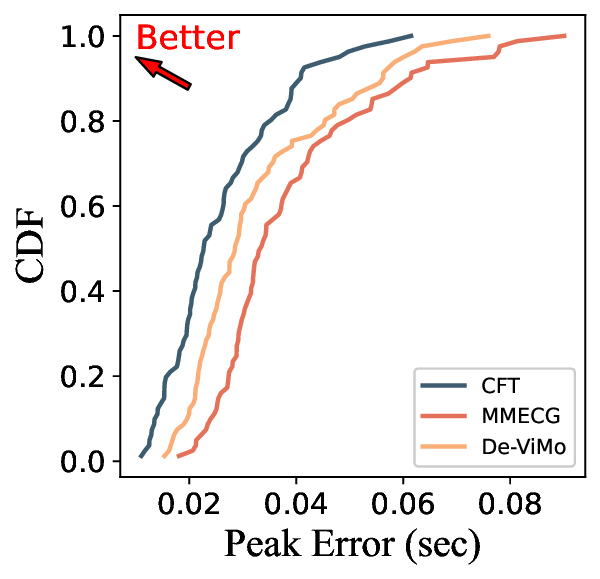}}
  \subfloat[]{\label{fig:pk_mdr_cdf}\includegraphics[width=0.5\columnwidth]{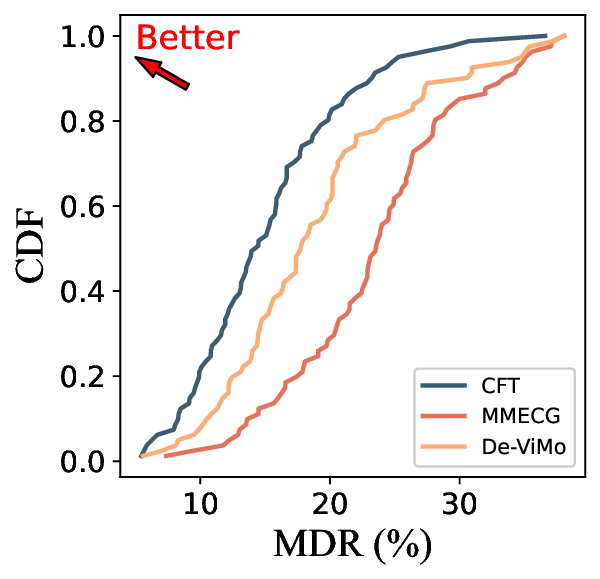}} 
  \caption{\textcolor{black}{Illustration of performance in terms of peak error and MDR from different perspectives: (a) - (b) Scattered points and fitting lines w.r.t monitoring range; (c) - (d) Scattered points and fitting lines w.r.t moving distance of the CF points; (e) - (f) CDF plots for all trials.}}
  \label{fig:range_impact}
\end{figure}

To evaluate the performance of different methods when increasing the monitoring range, the quality of extracted radar signals for all trials is evaluated in terms of peak error and MDR. Figure~\subref*{fig:pk_err_point} and~\subref*{fig:pk_mdr_point} illustrate the peak error and MDR for all $80$ trials, with fitting lines indicating the mean peak error or MDR. 

All the methods show similar performance in the short range and experience certain degradation with respect to the increasing range, because the SNR naturally decreases as the monitoring range increases. Following the link budget analysis for a similar radar platform used in~\cite{zhang2024radarODE}, the degradation in SNR is around $14$~dB when the monitoring range increasing from $0.5$m to $1.1$m, causing a similar trend of degradation as shown in Figure~\subref*{fig:pk_err_point} and~\subref*{fig:pk_mdr_point}.

In particular, MMECG shows larger degradation and variance for longer-range cases because the rough localization based on FMCW signal processing cannot provide accurate cardiac location, and the resultant evaluated points may not capture useful information for clustering. In contrast, De-ViMo could provide a better cardiac location, and the accumulated results show better accuracy than MMECG, while the long-range monitoring still affects the quality because the accumulation is not robust to non-Gaussian noise. At last, the proposed CFT could precisely focus on the CF point regardless of the monitoring range, providing the best results with small variance for both peak error and MDR.

\subsubsection{Impact of Posture Change}
As one of the core innovations, CFT could locate and keep tracking the CF point to ensure a high-quality signal acquisition after the posture change. To provide a quantified evaluation, the posture change is counted as the travel distance of the CF points during each $1$-minute trial to measure the extent of upper body movement, and the results are shown in Figure~\subref*{fig:pk_err_point_dis} and~\subref*{fig:pk_mdr_point_dis}. 

Similar to the impact of range, the peak error and MDR increase with moving distance because the signal collected during the posture change is heavily ruined, also known as random body movement distortion~\cite{chen2022contactless}. MMECG and De-ViMo will be affected if the CF points deviate too much from the initial body location and cannot guarantee a high SNR, because they do not have enough features for clustering or accumulation. In contrast, CFT could effectively find the new CF points after rounds of spatial search, and the extracted signal quality is better than the other two methods, as indicated by the fitting lines in Figure~\subref*{fig:pk_err_point_dis} and~\subref*{fig:pk_mdr_point_dis}. 

\subsubsection{Signal Qualities for All Trials}
The cumulative distribution function (CDF) plots for all trials are shown in Figure~\subref*{fig:pk_err_cdf} and~\subref*{fig:pk_mdr_cdf}. The proposed CFT algorithm achieves the best peak error with a median value of $0.022$~sec, while DE-ViMo and MMECG have worse performances with larger median values of $0.028$~sec and $0.033$~sec, respectively. Similarly, the precise localization and tracking of CF point also reduces the MDR for CFT results with a median value of $14\%$, while DE-ViMo and MMECG may be affected by the accumulated noise or inaccurate cardiac localization with the median MDR of $17\%$ and $23\%$, respectively.

\subsubsection{Evaluations of Computational Cost}
\begin{figure}[tb] 
    \centering 
    \subfloat[]{\label{fig:evaluation_num}\includegraphics[width=0.7\columnwidth]{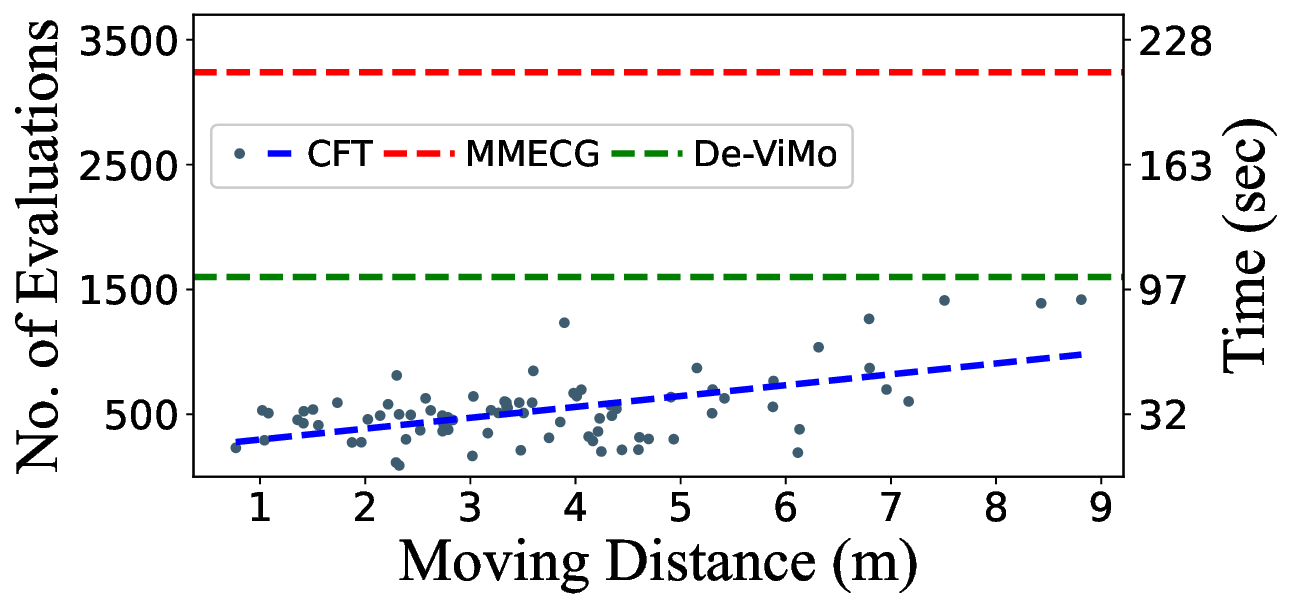}}\\
    \subfloat[]{\label{fig:evaluation_num_rag}\includegraphics[width=0.7\columnwidth]{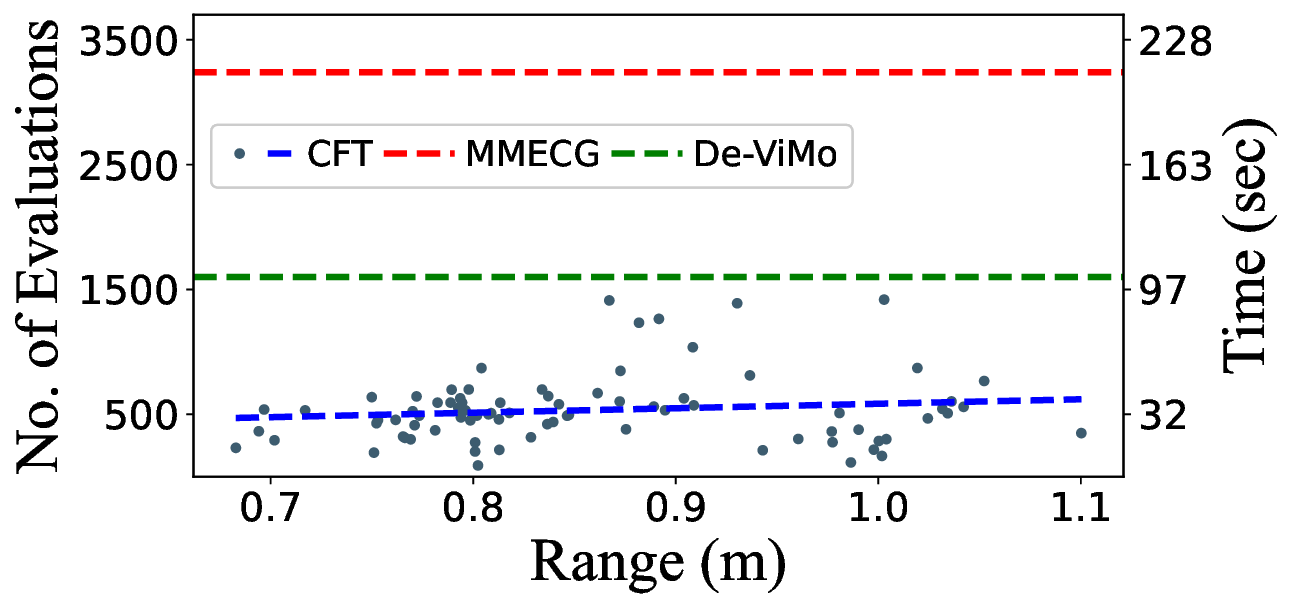}}
    \caption{\textcolor{black}{Number of evaluations and overall time consumed for each trial w.r.t the moving distance and monitoring range.}}
    \label{fig:move_dis} 
\end{figure}

The outstanding performance of CFT owes to the space search mechanism that triggers new rounds of search if the current points no longer satisfy the desired SNR (SNR$_d$). In comparison, MMECG and De-ViMo perform a fixed number of evaluations for every iteration, wasting a huge amount of time to calculate redundant points or channels. For all three methods, the most time-consuming step is to extract signals from points as calculated in (\ref{equ:raw_sig}) and (\ref{equ:phase}), while the time consumed for clustering, accumulation, and space search is negligible. Therefore, the quantitative computational cost can be estimated by counting the evaluation times for each 4-sec signal segment:
\begin{itemize}
  \item MMECG takes $216$ spatial points for clustering, and the total number of evaluations for a $1$-minute trial is $3240$.
  \item De-ViMo needs to consider the signals extracted from $2$ chirps, $8$ virtual antennas, $10$ spatial points, and the number of evaluations is $2400$.
  \item CFT only evaluates the point during space search to get the cost values, and the number of evaluations recorded for each trial is shown in Figure~\ref{fig:move_dis}.
\end{itemize}

According to the experiment, each evaluation takes about $65$ms, and the overall time consumed for evaluating a $1$-minute trial is also shown in Figure~\ref{fig:move_dis}. Apparently, the proposed CFT saves a huge amount of time compared with other methods, and the time used for space search increases if the subjects change their posture more frequently (i.e., with longer moving distance), as shown in Figure~\subref*{fig:evaluation_num_rag}. In addition, there is no obvious relationship shown between evaluation numbers and monitoring ranges as shown in Figure~\subref*{fig:evaluation_num}, indicating that the CFT algorithm can keep extracting high-quality radar signals regardless of monitoring range and could save huge computational resources if the subject is static.
\color{black}

\subsubsection{Impact of Signal Quality on ECG Recovery}
\begin{figure}[tb]
  \centering
  \subfloat[]{\label{fig:radar_clean}\includegraphics[width=0.5\columnwidth]{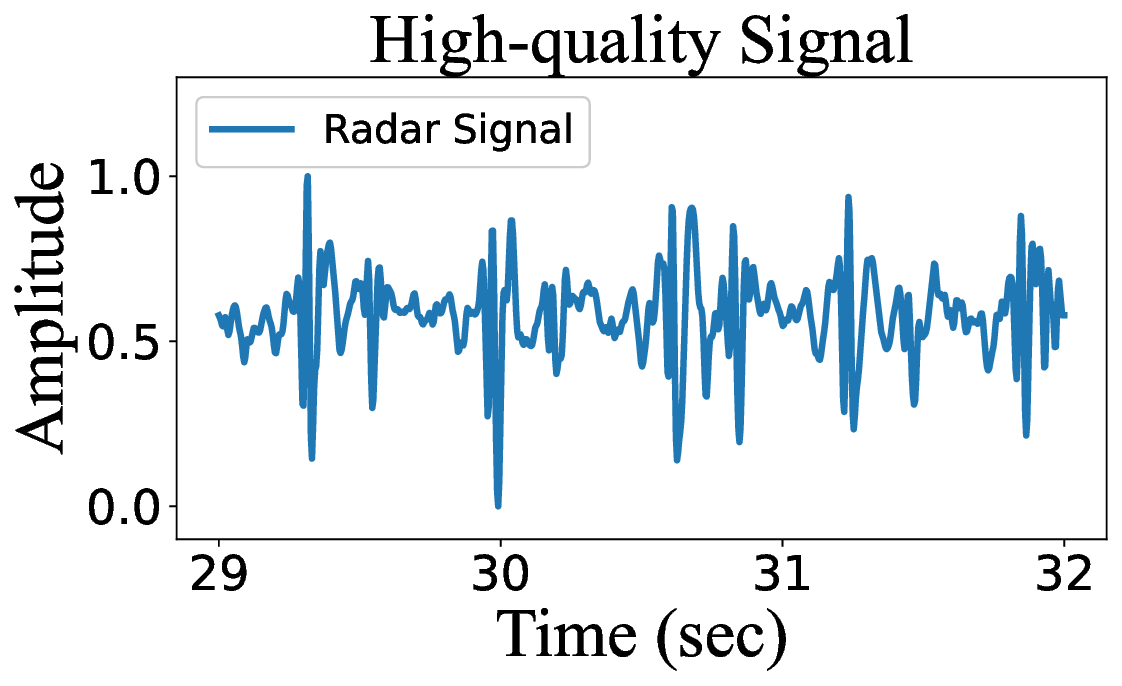}}
  \subfloat[]{\label{fig:ecg_pred_clean}\includegraphics[width=0.5\columnwidth]{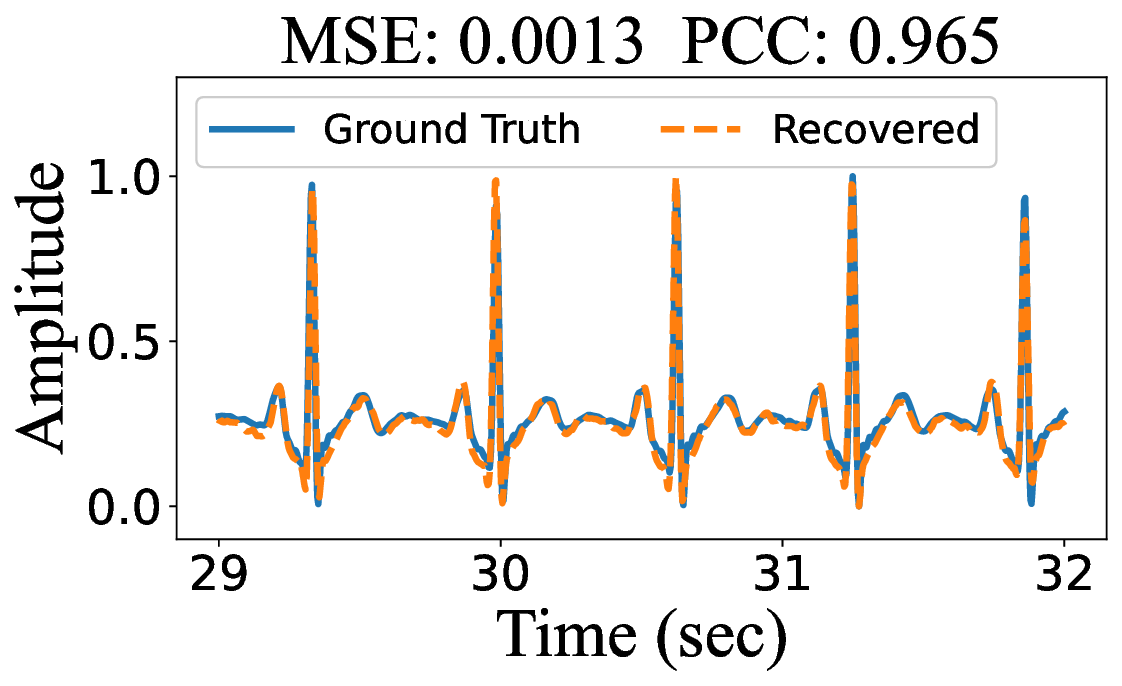}} \\
  \subfloat[]{\label{fig:radar_noise}\includegraphics[width=0.5\columnwidth]{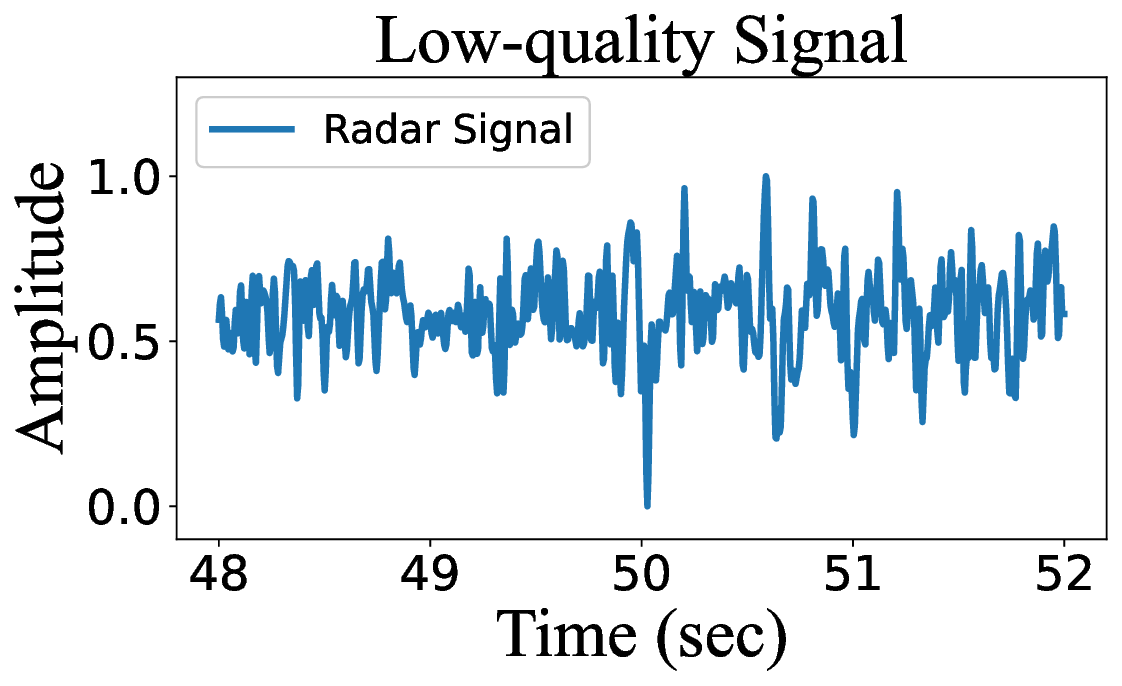}}
  \subfloat[]{\label{fig:ecg_pred_noise}\includegraphics[width=0.5\columnwidth]{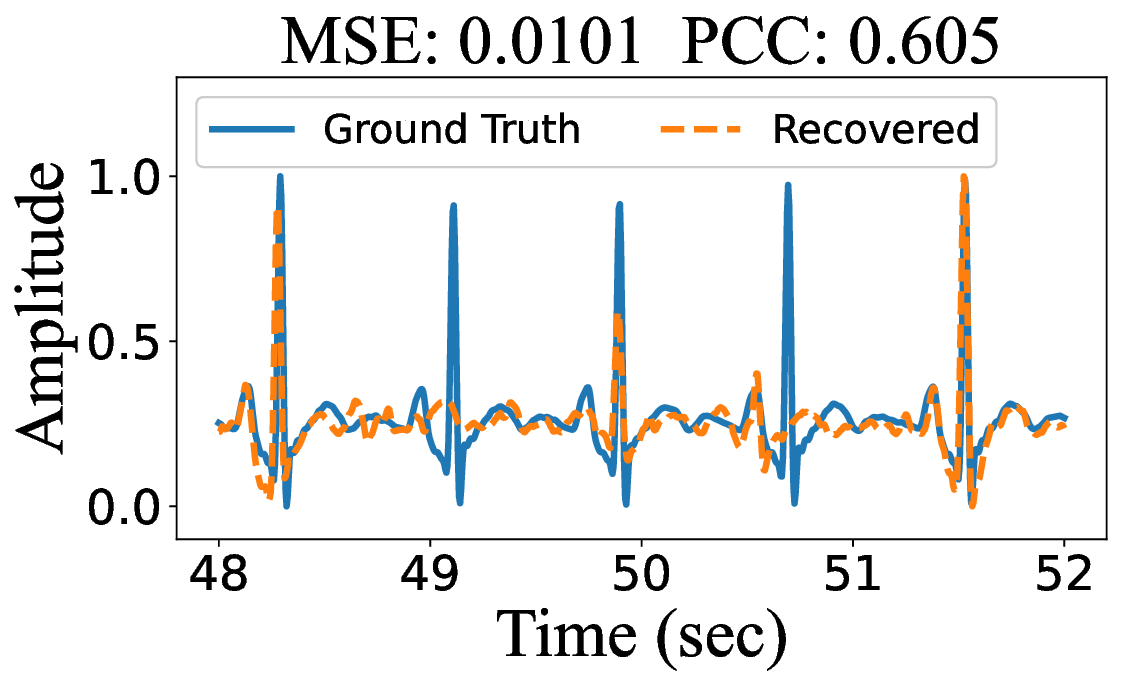}} 
  \caption{Impact of radar input quality on the final ECG recovery: (a) - (b) High-quality radar input and ECG recovery; (c) - (d) Low-quality radar input and ECG recovery.}
  \label{fig:ecg_recovery}
\end{figure}

The signals extracted using different methods are used for supervised training to verify the impact of different input qualities on the ECG recovery task. Table~\ref{tab:ecg_supervise} shows the performance of the deep learning model trained with datasets yielded by different methods. The training based on CFT dataset achieves the best results on both morphological accuracy (MSE$=0.0082$ and PCC$=85.47\%$) and R peak recovery (Peak Error$=7.61$ms and MDR$=6.85\%$), because the high-SNR inputs provide accurate peak locations with minor noise that affects the ECG pattern generation, as shown in Figure~\subref*{fig:radar_clean} and~\subref*{fig:ecg_pred_clean}. 

In contrast, MMECG and De-ViMo cannot preserve the signal quality especially for long-distance cases, and the noisy inputs will prevent the deep learning model from identifying the accurate position of ECG pieces, causing large peak error and MDR, as shown in Figure~\subref*{fig:radar_noise} and~\subref*{fig:ecg_pred_noise}. It is worth noticing that poor signal SNR causes more degradation in peak error than morphological accuracy, because the ECG patterns share a similar shape and can be learned from other cardiac cycles, while the peak recovery (detection) fully relies on the current radar input and can be ruined by noises.

\begin{table}[tb]
\centering
\caption{Performance of supervised ECG recovery}
    \begin{tabular}{c |cccc}
    \toprule
    Methods  & \makecell[c]{MSE\\($\times 10^{-2}$)} $\downarrow$ & PCC $\uparrow$ & \makecell[c]{Peak Error \\ (ms)} $\downarrow$ & MDR $\downarrow$ \\
    \toprule
    MMECG~\cite{chen2022contactless} & $0.93$  & $80.36\%$ & $9.74$ & $7.96\%$ \\
    De-ViMo~\cite{liu2024diversity} & $0.88$ & $83.83\%$ & $8.93$ & $7.32\%$\\
    \midrule
    CFT  &  $\mathbf{0.80}$ & $\mathbf{85.47\%}$ & $\mathbf{7.61}$ & $\mathbf{6.85\%}$\\
    \bottomrule
    \end{tabular}%
\label{tab:ecg_supervise}
\end{table}%

\begin{figure}[tb]
  \centering
  \subfloat[]{\label{fig:ssr_bad}\includegraphics[width=0.5\columnwidth]{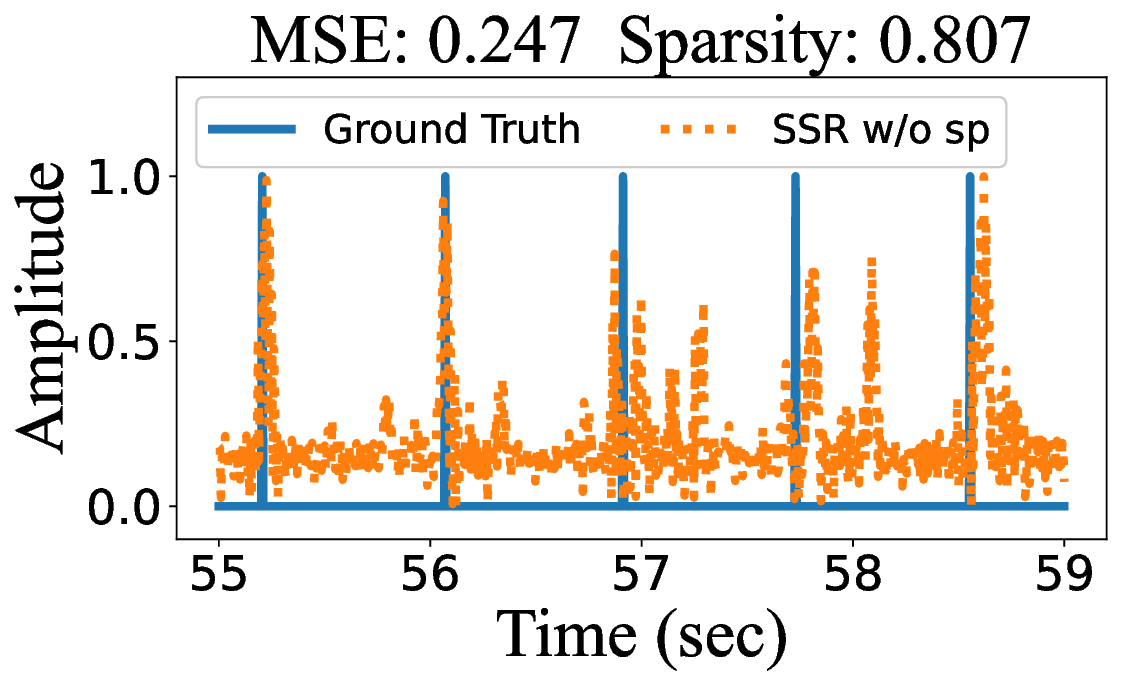}}
  \subfloat[]{\label{fig:ssr_good}\includegraphics[width=0.5\columnwidth]{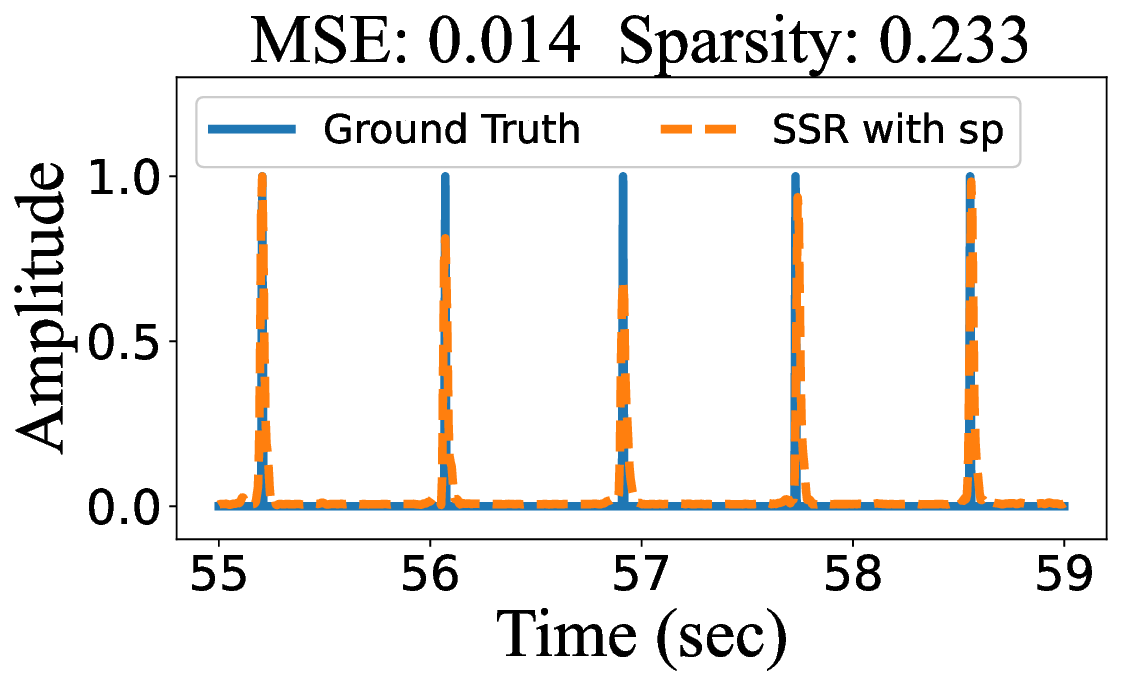}}
  \caption{Results of SSR: (a) Failed SSR due to lack of data and sparse penalty; (b) Ideal SSR result with good MSE and sparsity.}
  \label{fig:ssr_res}
\end{figure}

\subsection{Performance of ECG Recovery using Transfer Learning}
\subsubsection{Evaluations and Ablation Studies of SSL in RFcardi}
The SSR task is crucial in the proposed transfer learning framework to provide latent representations that assist further ECG pattern recovery, reducing the demand for radar-ECG pairs in the fine-tuning stage. The results of SSL are shown in Table~\ref{tab:ssr} in terms of MSE and sparsity to illustrate the former (MSE term) and latter part (sparse penalty $\lambda_s$) in the loss function (\ref{equ:sparse}) for SSL training. 

The experiment is repeated for different dataset scales with the ablation study on the use of sparse penalty, and the results indicate that both MSE and sparsity decrease with the reduction in training data as shown in Table~\ref{tab:ssr}. Training with $100\%$ or $80\%$ dataset could achieve convergence and realize a successful SSR with similar MSE ($0.0091, 0.0082, 0.0081$) or ($0.0096, 0.0085, 0.0088$). However, the performance of SSR degrades heavily when further decreasing the training data without the constraint of sparse penalty ($\lambda_s=0$), because the SSR results might fluctuate, as shown in Figure~\subref*{fig:ssr_bad}. In contrast, introducing the sparse penalty could suppress the fluctuation and force the deep learning model to focus only on the dominant peaks of the input radar signals, as shown in Figure~\subref*{fig:ssr_good}. Therefore, the training with sparse penalty loss still achieves good results with an MSE of $0.0092$ and $0.0098$ by using $60\%$ and $40\%$ of the dataset, while the model cannot be well-trained without sparse penalty by using $40\%$ dataset as shown in Table~\ref{tab:ssr} and Figure~\subref*{fig:ssr_bad}.

\color{black}
It is also revealed in Table~\ref{tab:ssr} about the consequence of enlarging the sparse penalty $\lambda_s$. Considering that even the ground truth shown in Figure~\ref{fig:ssr_res} has a sparsity of $0.05$, using a large coefficient (e.g., $\lambda_s=0.05$) will overwhelm the MSE term, causing the DL model to generate the result with only $1$ non-zero value to give a sparsity of $0$. In Table~\ref{tab:ssr}, $\lambda_s=0.02$ gives a higher MSE with lower sparsity compared with $\lambda_s=0.01$, especially when using $60\%$ and $40\%$ of the dataset (MSE$=0.0118$ and $0.0124$, sparsity$=0.024$ and $0.026$). In other words, using a large sparse penalty cannot help the convergence when using a limited dataset, but may disrupt the training process due to the dominance of the sparse penalty term, causing suboptimal performance in capturing cardiac-related features (i.e., heartbeats).

\begin{table*}[tb]
\centering
\color{black}
\caption{Performance of SSL and fine-tuning with ablation study using different $\lambda_s$}
    \begin{tabular}{l |cc|cccc|c}
    \toprule
    \makecell[c]{\multirow{2}*{Methods}} &  \multicolumn{2}{c|}{\textbf{SSL}} & \multicolumn{4}{c|}{\textbf{Fine-tuning}} & \multirow{2}*{Overall $\uparrow$} \\
    & \makecell[c]{MSE ($\times 10^2$)} $\downarrow$ & Sparsity $\downarrow$ & \makecell[c]{MSE ($\times 10^{-2}$)} $\downarrow$ & PCC $\uparrow$ & \makecell[c]{Peak Error (ms)} $\downarrow$ & MDR $\downarrow$  &\\
    \toprule
    & \multicolumn{2}{c|}{\textbf{$\mathbf{100\%}$ Unlabeled}} & \multicolumn{4}{c|}{\textbf{$\mathbf{100\%}$ Labeled}} \\ 
    \midrule
    RFcardi with $\lambda_s=0$ & ${0.91}$ & ${0.36}$ & $0.81$ & $85.35\%$ & $8.46$ & $5.51\%$ & $1.75\%^1$ \\
    RFcardi with $\lambda_s=0.01$ & ${0.82}$ & ${0.20}$ & $0.80$ & $85.51\%$ & $8.40$ & $5.14\%$ & $\mathbf{3.66\%}$ \\
    RFcardi with $\lambda_s=0.02$ & ${0.81}$ & ${0.14}$ & $0.80$ & $86.60\%$ & $8.34$ & $5.47\%$ & ${2.97\%}$\\
    \midrule
        & \multicolumn{2}{c|}{\textbf{$\mathbf{80\%}$ Unlabeled}} & \multicolumn{4}{c|}{\textbf{$\mathbf{80\%}$ Labeled}} \\ 
    \midrule
    RFcardi with $\lambda_s=0$    & $0.96$ & $0.41$ & $0.88$ & $82.69\%$ & $9.97$ & $7.15\%$ & $-12.18\%$ \\
    RFcardi with $\lambda_s=0.01$ & $0.85$ & $0.22$ & $0.82$ & $84.53\%$ & $9.95$ & $6.94\%$ & $\mathbf{-8.92\%}$\\
    RFcardi with $\lambda_s=0.02$ & $0.88$ & $0.17$ & $0.83$ & $83.23\%$ & $8.71$ & $8.17\%$ & $-10.02\%$ \\
    \midrule
        & \multicolumn{2}{c|}{\textbf{$\mathbf{60\%}$ Unlabeled}} & \multicolumn{4}{c|}{\textbf{$\mathbf{60\%}$ Labeled}} \\ 
    \midrule
    RFcardi with $\lambda_s=0$    & $1.43$ & $0.44$ & $0.90$ & $80.45\%$ & $9.68$ & $9.44\%$ & $-20.85\%$ \\
    RFcardi with $\lambda_s=0.01$ & $0.92$  & $0.26$ & $0.91$ & $81.56\%$ & $9.31$ & $7.58\%$ & $\mathbf{-12.83\%}$ \\
    RFcardi with $\lambda_s=0.02$ & $1.18$ & $0.24$  & $0.94$ & $82.88\%$ & $8.17$ & $9.7\%$ & $-17.37\%$  \\
    \midrule
        & \multicolumn{2}{c|}{\textbf{$\mathbf{40\%}$ Unlabeled}} & \multicolumn{4}{c|}{\textbf{$\mathbf{40\%}$ Labeled}} \\ 
    \midrule
    RFcardi with $\lambda_s=0$ &  - & - & - & -& -&-& Failed$^2$ \\
    RFcardi with $\lambda_s=0.01$ & $0.98$  & $0.31$ & $0.99$ & $76.87\%$ & $11.09$ & $8.85\%$ & $\mathbf{-27.18\%}$  \\
    RFcardi with $\lambda_s=0.02$ & $1.24$  & $0.26$ & $1.13$ & $66.43\%$ & $12.23$ & $10.41\%$ & $-44.05\%$\\
    \bottomrule
    \multicolumn{8}{l}{$^1$The benchmark with $0\%$ improvement is given in Table~\ref{tab:ecg_few_shot} for supervised learning.} \\
    \multicolumn{8}{l}{$^2$The ECG recovery fails if $PCC< 60\%$, according to the empirical observation of the morphological ECG features.}
    \end{tabular}%
\label{tab:ssr}
\end{table*}%
\subsubsection{Evaluations and Ablation Studies of Fine-tuning Results}\label{sec:ssl_eval}
\begin{figure}[tb]
  \centering
  \subfloat[]{\label{fig:rf_overall}\includegraphics[width=0.5\columnwidth]{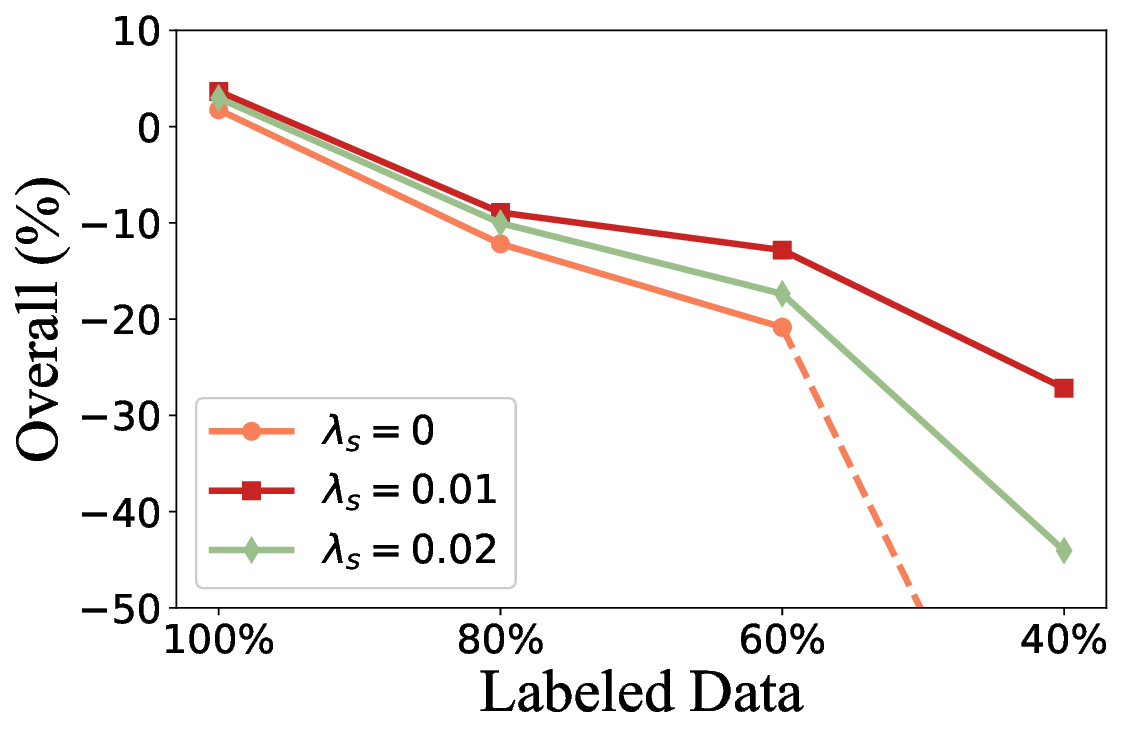}}
  \subfloat[]{\label{fig:overall}\includegraphics[width=0.5\columnwidth]{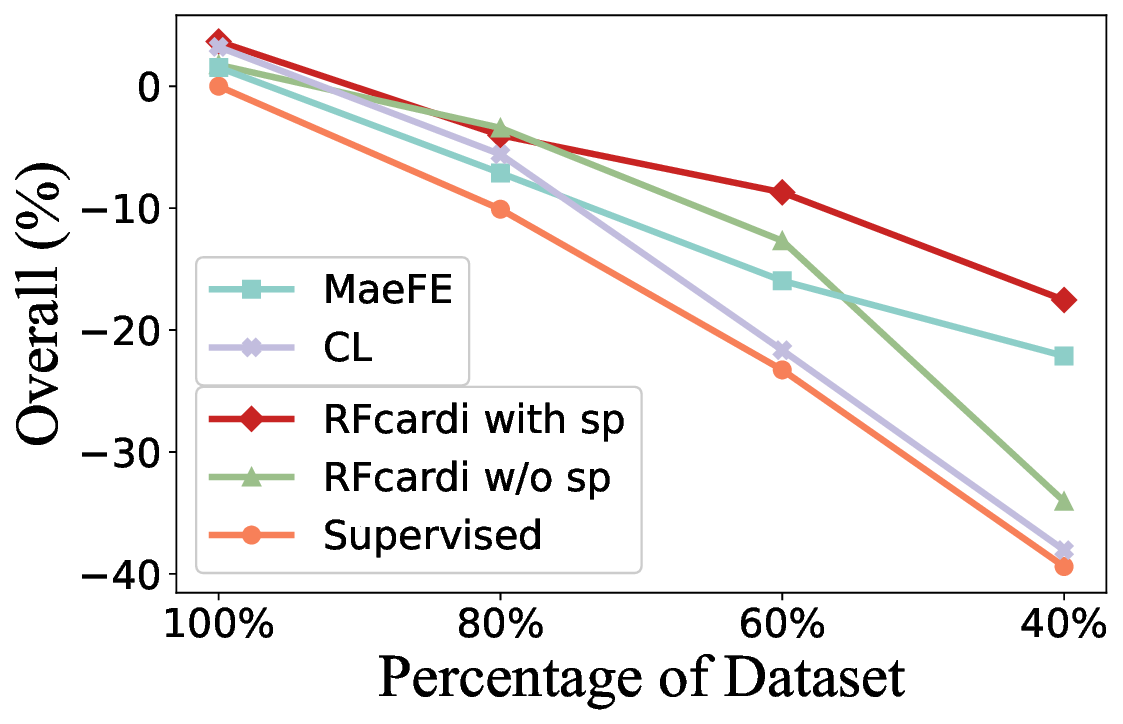}}
  \caption{\textcolor{black}{Overall performance of the radar-based ECG recovery: (a) RFcardi with different $\lambda_s$; (b) Comparison with different SSL methods.}}
  \label{fig:overall_two}
\end{figure}

To verify the impact of SSL quality on the fine-tuning result, the same amount of labeled data is used for fine-tuning with different sparse penalty $\lambda_s$, as shown in Table~\ref{tab:ssr}. The overall improvements are also provided in the last column by calculating the percentage of improvement across all four metrics to provide straightforward evaluations for different methods, as shown in Table~\ref{tab:ssr} and Figure~\subref*{fig:rf_overall}.

The results coincide with the expectation that more labeled data generates better ECG recovery performance, and the discrepancy between using different $\lambda_s$ is not obvious if the training data is enough (e.g., for $100\%$ and $80\%$), as indicated in Figure~\subref*{fig:rf_overall}. However, if the labeled data is not enough, the best pre-training model ($\lambda_s=0.01$) gives the best ECG recovery for all cases with different labeled data scales, because SSL with $\lambda_s=0.02$ or $0$ does not properly train the backbone with the ability to capture the peak-related information. In addition, the gap between using different $\lambda_s$ is larger than the case with enough data as shown in Figure~\subref*{fig:rf_overall}, revealing the advantage of using SSL for pre-training. According to our experiment, $\lambda_s=0.01$ guarantees a good performance for different scales of input datasets and will be adopted for the comparison with other SSL methods later.

\begin{table*}[tb]
\centering
\color{black}
\caption{Performance of ECG recovery using different percentages of labeled data}
    \begin{tabular}{l |cccc|c?cccc|c}
    \toprule
    \makecell[c]{Methods} & \makecell[c]{MSE\\($\times 10^{-2}$)} $\downarrow$ & PCC $\uparrow$ & \makecell[c]{Peak \\ Error \\ (ms)} $\downarrow$ & MDR $\downarrow$ & Overall $\uparrow$ & \makecell[c]{MSE\\($\times 10^{-2}$)} $\downarrow$ & PCC $\uparrow$ & \makecell[c]{Peak \\ Error \\ (ms)} $\downarrow$ & MDR $\downarrow$  & Overall $\uparrow$\\
    \toprule
    &\multicolumn{4}{c|}{\textbf{$\mathbf{100\%}$ Unlabeled, $\mathbf{100\%}$ Labeled}} && \multicolumn{4}{c|}{\textbf{$\mathbf{100\%}$ Unlabeled, $\mathbf{80\%}$ Labeled}} \\ 
    \midrule
    Supervised & ${0.80}$ & ${85.47\%}$ & ${7.61}$ & ${6.85\%}$ & $0.00\%$ & $0.84$ & $84.60\%$ & $8.90$ & $8.04\%$ & $-10.09\%$ \\
    MaeFE~\cite{zhang2022maefe} & $0.81$ & $83.97\%$ & $7.98$ & $5.89\%$ & $1.54\%$ & $0.84$ & $83.61\%$ & $9.22$ & $6.86\%$ & $-7.12\%$\\
    CL~\cite{canellas2025self} & $0.81$ & $86.15\%$ & $8.85$ & $4.47\%$ & ${3.22\%}$ & $0.82$ & $84.30\%$ & $9.96$ & $6.00\%$ & $-5.59\%$ \\
    {RFcardi w/o sp$^*$} & $0.81$ & $85.35\%$ & $8.46$ & $5.51\%$ & $1.75\%$ & $0.82$ & $86.36\%$ & $8.35$ & $7.02\%$ & $\mathbf{-3.42\%}$ \\
    {RFcardi with sp} & $0.80$ & $85.51\%$ & $8.40$ & $5.14\%$ & $\mathbf{3.66\%}$ & $0.81$ & $84.29\%$ & $8.31$ & $7.14\%$ & ${-4.02\%}$ \\
    \midrule
       & \multicolumn{4}{c|}{\textbf{$\mathbf{100\%}$ Unlabeled, $\mathbf{60\%}$ Labeled}}& & \multicolumn{4}{c|}{\textbf{$\mathbf{100\%}$ Unlabeled, $\mathbf{40\%}$ Labeled}} \\ 
    \midrule
    Supervised & $0.93$ & ${79.91\%}$ & ${10.65}$ & ${8.93\%}$ & $-23.27\%$ & $0.98$ & $75.89\%$& $11.15$ & $12.15\%$ & $-39.40\%$\\
    MaeFE~\cite{zhang2022maefe} & $0.87$ & $82.01\%$ & $9.46$ & $8.68\%$ & $-15.96\%$ & $0.96$ & $78.51\%$ & $10.36$ & $8.51\%$ & ${-22.13\%}$ \\
    CL~\cite{canellas2025self} & $0.92$ & $78.83\%$ & $10.12$ & $12.42\%$ & $-21.65\%$ & $1.05$ & $75.39\%$ & $11.03$ & $11.25\%$ & $-38.05\%$\\
    {RFcardi w/o sp} & $0.85$ & $83.74\%$ & $8.84$ & $8.65\%$ & ${-12.68\%}$ & $0.97$ & $76.56\%$ & $10.87$ & $10.99\%$ & $-34.05\%$ \\
    {RFcardi with sp} & $0.86 $& $84.92\%$ & $8.58$ & $7.72\%$ & $\mathbf{-8.71\%}$ & $0.93$ & $78.72\%$ & $8.70$ & $9.02\%$ & $\mathbf{-17.54\%}$ \\
    \bottomrule
    \multicolumn{9}{l}{$^*$sp for sparse penalty ($\lambda_s=0.01$)}
    \end{tabular}%
\label{tab:ecg_few_shot}
\end{table*}%
\subsubsection{Comparison with Other SSL Methods}
The comparison is based on the model pre-trained on a $100\%$ dataset using different SSL methods, and the experiment is repeated for different percentages of labeled data (i.e., radar signal with ECG ground truth). In addition, the same deep-learning model will be trained in a supervised manner by using the same amount of labeled data as the benchmark for transfer-learning ECG recovery. The overall performances are also listed in the last column and plotted in Figure~\subref*{fig:overall}.

The fine-tuning with $100\%$ labeled data provides very similar performance in morphological accuracy, with the MSE and PCC around $0.0080$ and $85.47\%$. It is worth noticing that the peak error and MDR are slightly improved, because the pre-text task SSR for pre-training is equivalent to identifying the peak position of the radar signal, and the learned representations can be seamlessly transferred to improve the accuracy of the recovered ECG R peaks, contributing to the overall improvement for RFcardi ($3.66\%$ and $1.75\%$). Similarly, MaeFE and CL also achieve slight improvement, although they are not pre-trained by the cardiac-feature-related pre-text tasks, enough data still guarantees an effective pre-training and fine-tuning.

Reducing $20\%$ of labeled data causes a $10\%$ overall degradation for supervised learning as shown in Table~\ref{tab:ecg_few_shot}, and the decline of peak error and MDR is more than MSE and PCC. The reason is that ECG morphological patterns for different cardiac cycles are similar and can be well-learned from $80\%$ labeled data with good MSE and PCC ($0.0084$ and $84.60\%$), while the location of each ECG piece is random and can be distorted by noise, requiring more training data for convergence. In addition, $80\%$ dataset is still enough for MaeFE and CL to achieve a similar performance with RFcardi, with the performance of $-7.12\%$ and $-5.19\%$ compared with $-4.02\%$ for RFcardi with sparse penalty.

The supervised training with $60\%$ labeled data cannot ensure a good morphological and peak accuracy, and the overall degradation is $23.37\%$, with the PCC drop below $80\%$ as shown in Table~\ref{tab:ecg_few_shot}. In contrast, the pre-trained model still provided good results with mild degradations of $8.71\%$ and $12.68\%$. It is noticed that the effectiveness of the sparse penalty in the SSL stage also affects the fine-tuning stage, because both peak error and MDR for transfer learning with sparse penalty are better than those without sparse penalty, causing a large gap in the overall improvement compared with the previous training with $100\%$ and $80\%$ datasets. In this stage, CL shows a larger gap compared with other SSL methods as shown in Figure~\subref*{fig:overall}, with the overall performance of $-21.65\%$, close to the result achieved by supervised learning ($23.27\%$). The reason is that CL relies on the distinction between positive and negative samples, while the construction designed in~\cite{canellas2025self} is not suitable for cardiac signals, because the cardiac activities may share the same pattern even after applying more than $20$ seconds of delay.

\begin{figure}[tb]
  \centering
  \subfloat[]{\label{fig:fs_bad}\includegraphics[width=0.5\columnwidth]{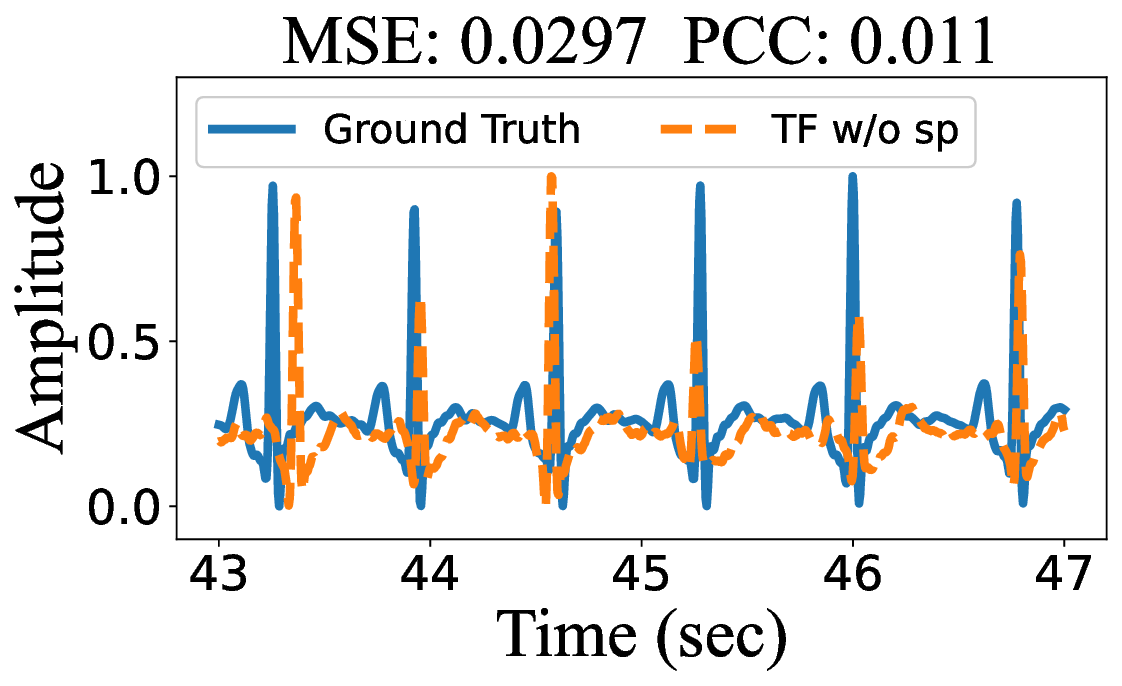}}
  \subfloat[]{\label{fig:fs_good}\includegraphics[width=0.5\columnwidth]{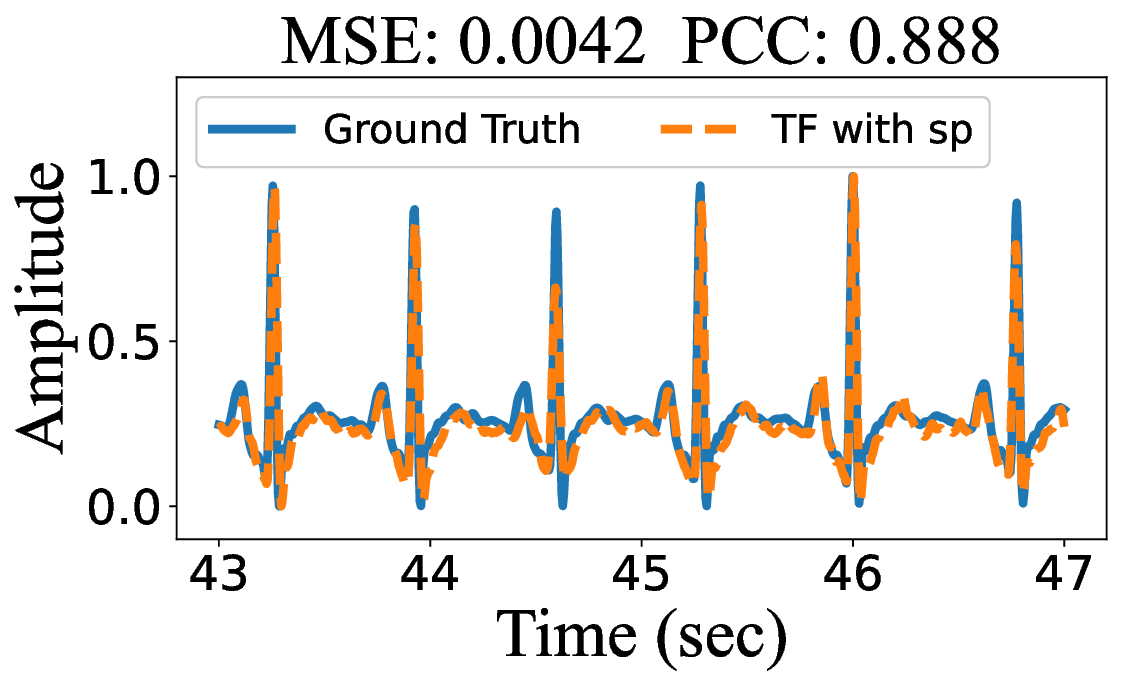}}
  \caption{Results of transfer learning using limited labeled data: (a) Poor ECG recovery without proper morphological feature and peak location; (b) Good ECG recovery owing to the well pre-trained model.}
  \label{fig:fs_res}
\end{figure}

Lastly, the deep learning model can barely learn from $40\%$ labeled dataset and yield a bad morphological and peak accuracy for supervised learning. In addition, CL, MaeFE, and RFcardi, without a sparse penalty, all struggle to learn both morphological and peak features from limited data if the pre-text task is not well-trained. CL and MaeFE do not leverage specially designed pre-text to highlight the peak-related features and cannot achieve an effective SSL with limited data, as evaluated in Section~\ref{sec:ssl_eval}. Similarly, RFcardi without a sparse penalty cannot achieve convergence during SSL, with the examples of the recovered ECG signals shown in Figure~\ref{fig:fs_res}. It is obvious that the well-pretrained model generates the ECG signal with good peak and morphological accuracy after fine-tuning as shown in Figure~\subref*{fig:fs_good}, while the large peak deviation and bad ECG morphology are caused by the pre-trained models that cannot capture essential peak-related features, as shown in Figure~\subref*{fig:fs_bad}.

\subsubsection{Summary of Transfer-learning-based ECG Recovery}
Previous evaluations in terms of SSL and fine-tuning stages have illustrated the ability of the proposed RFcardi to learn from unlabeled data and transfer the knowledge to the ECG recovery task using limited radar-ECG pairs. Here are some key conclusions drawn from Figure~\ref{fig:overall_two}:
\begin{itemize}
  \item The performance of supervised learning drops heavily and cannot ensure high-quality ECG recovery after reducing $40\%$ labeled dataset.
  \item Transfer learning could enhance the performance of ECG recovery for cases with ample labeled training data, and the quality of the pre-trained model has a minor effect on the final result because the deep learning model could learn from numerous radar-ECG pairs.
  \item The proposed RFcardi shows outstanding performance in learning latent representations from unlabeled data through a specially designed pre-text task, SSR. In comparison, MaeFE and CL are not designed for this task and cannot effectively capture the peak-related features when the input data is limited. 
  \item For the cases with limited labeled data ($40\%$, $60\%$), the quality of the pre-trained model does matter to alleviate the burden of the deep learning model to learn both morphological ECG patterns and peak locations, as indicated by the increasing gap in both Figure~\subref*{fig:rf_overall} and~\subref*{fig:overall}.
\end{itemize}

\subsection{Limitations and Future Work}
\subsubsection{Real-time Processing}
Figure~\ref{fig:move_dis} indicates that the time consumed for each $1$-minute trial is around $30$~sec, but the proposed CFT algorithm still needs improvement to achieve real-time processing, because the searching time can reach $6.5$ sec ($k_{max}=100$) for a $4$-sec segment if the collected raw radar signal is heavily distorted or the CF point is far from the current $E_k$. In future work, real-time processing ability should be considered as an essential ability, and the high-SNR signal collection algorithm should be tested on the edge devices to enable real-life applications.

\subsubsection{Energy Consumption}
In this study, the CFT algorithm requires multiple function evaluations, and the RFcardi model involves matrix multiplications and convolutional operations that are computationally intensive during inference. Considering that the mobile platforms typically face strict power budgets, the searching strategy of CFT can be optimized to reduce point evaluations by using adaptive grid adjustment and search directions~\cite{chu2025vessel2}. For deep learning, model pruning and knowledge distillation can decrease the computational load, reducing both runtime and energy demand during inference~\cite{zhao2025repair}.

\subsubsection{Multi-target Vital Sign Monitoring}
In real-world monitoring, a multi-target scenario will introduce extra interference and also require the separation of the signals reflected from different subjects~\cite{zhang2023overview}. The future work should focus on high-resolution and adaptive beamforming to better isolate the signal reflected from individual targets, and the CFT algorithm should be advanced to support multi-target vital signal extraction.

\color{black}

\section{Conclusions}\label{sec:conclusions}
This paper investigates the efficient collection of high-SNR radar signals with ample cardiac features for transfer-learning-based ECG recovery. Previous methods adopted signal accumulation or clustering to suppress the noise, while the rough localization based on FMCW radar cannot accurately reveal the chest region, requiring a time-consuming traverse through a 3D space for compensation. In this paper, a novel CFT algorithm is proposed to dynamically find the points with the best SNR and could track the cardiac location over time if the subjects change posture. In addition, a transfer learning framework RFcardi is designed with SSR as a pre-text task for pre-training to reduce the dependency on cumbersome ECG ground truth collection. The experiments performed in different scenarios prove the feasibility of the CFT-RFcardi framework in radar signal extraction and ECG recovery with limited labeled data, enabling a convenient deployment in new scenarios with limited data for remote healthcare monitoring.


\begin{IEEEbiography}[{\includegraphics[width=1in,height=1.25in,clip,keepaspectratio]{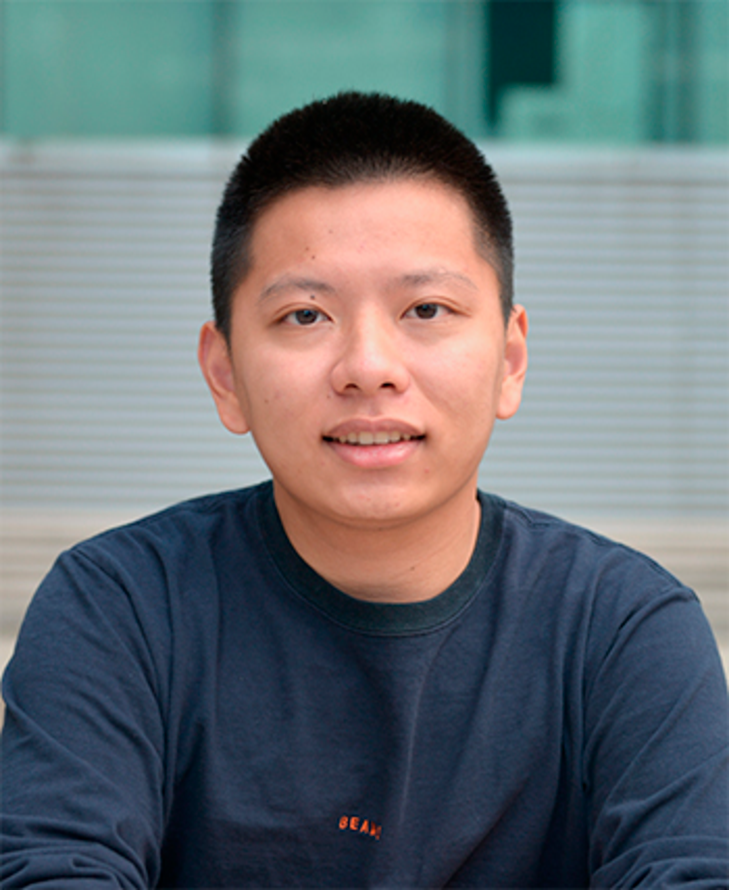}}]{Yuanyuan Zhang}
received the B.Eng. degree in electrical and electronic engineering from the University of Liverpool, UK, in 2020. He received M.S. degree in control system from the Imperial College London, UK, in 2021. He is currently pursuing the Ph.D. degree at the University of Liverpool, UK. His current research interests include wireless sensing, multi-task learning, optimization and sparse signal processing. 
\end{IEEEbiography}

\begin{IEEEbiography}[{\includegraphics[width=1in,height=1.25in,clip,keepaspectratio]{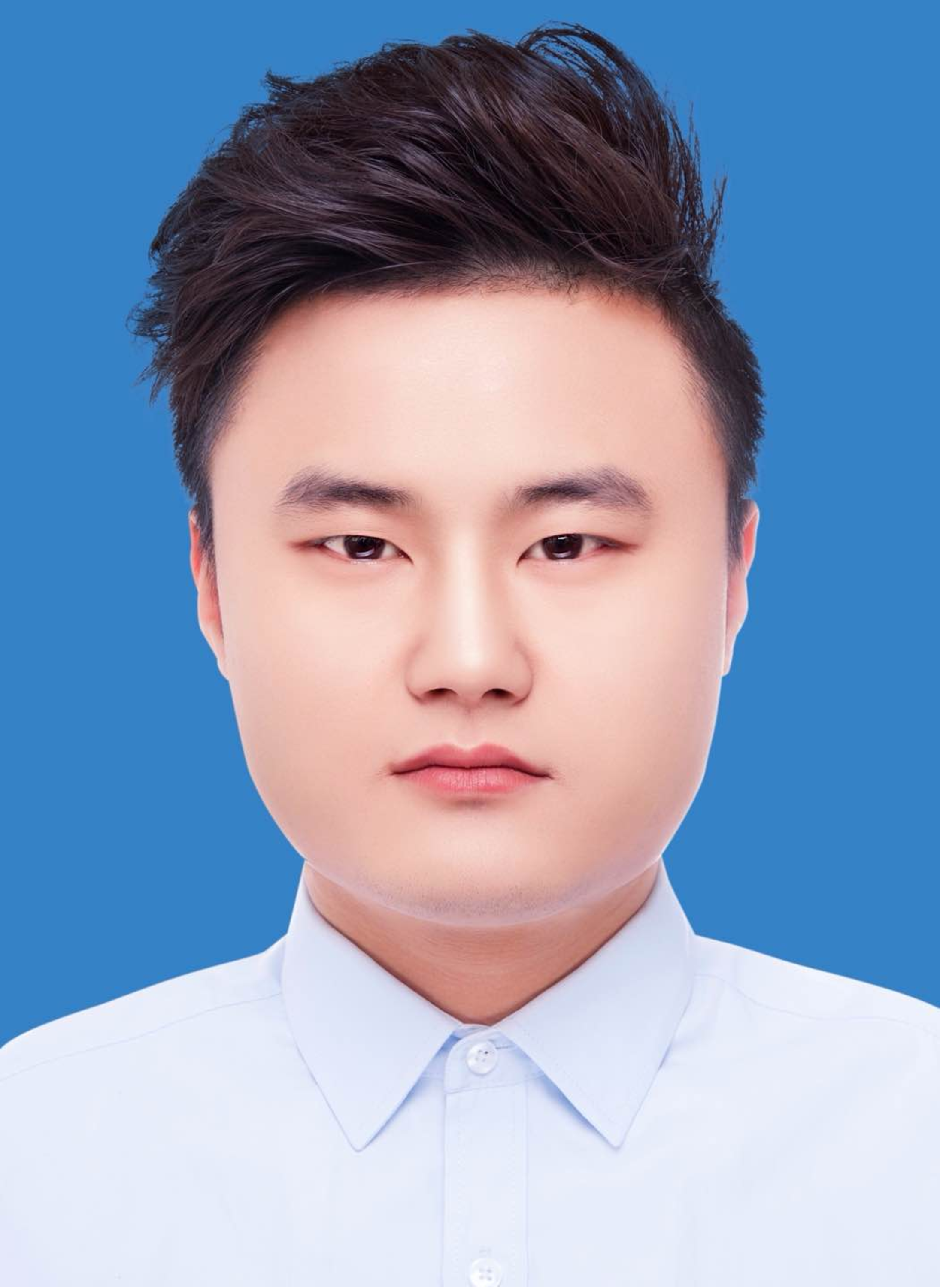}}]{Haocheng Zhao} received his B.E. and M.S. degree in 2019 and 2021, respectively, at Xi'an Jiaotong-Liverpool University, Jiangsu, China. He is currently a joint Ph.D. student of University of Liverpool, Xi'an Jiaotong-Liverpool University, Institute of Deep Perception Technology, and Jiangsu Industrial Technology Research Institute. His research interests include neural network pruning, radar perception, and robotics.
\end{IEEEbiography}

\begin{IEEEbiography}[{\includegraphics[width=1in,height=1.25in,clip,keepaspectratio]{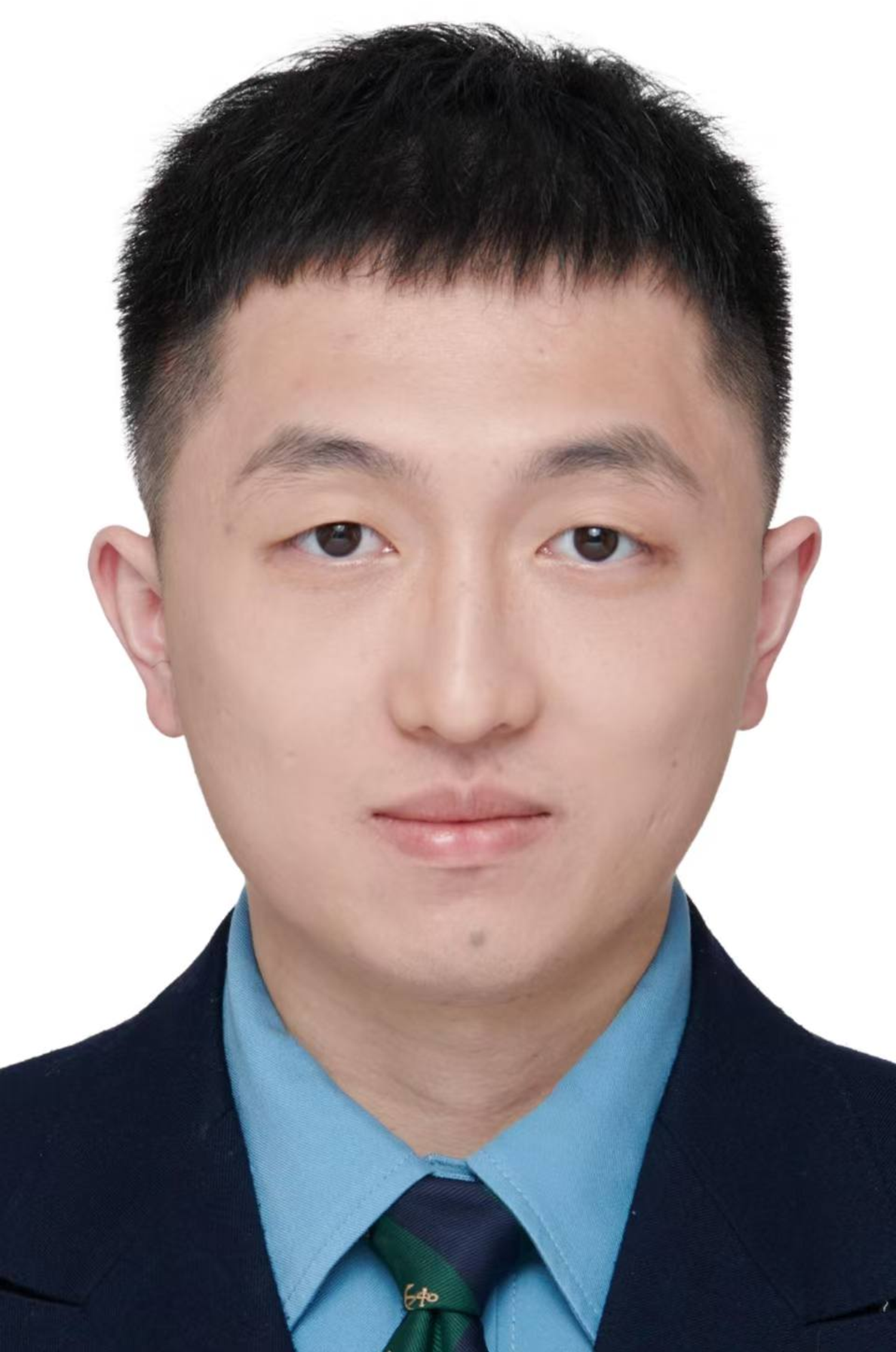}}]{Sijie Xiong} received the B.Eng. Degree in Automation from Hohai University, China, in 2020. He also received B.Sc. (Honor) Degree in Applied Accounting from Oxford Brookes University jointly with ACCA, UK, 2021. He received M.Sc. degree in Control Systems from Imperial College London, UK, 2021. He is currently pursuing the Ph.D. degree at Kyushu University, Japan. His current research interests include control theory, model order reduction, vibration reduction, and applications of AI in medical monitoring systems.
\end{IEEEbiography}

\begin{IEEEbiography}[{\includegraphics[width=1in,height=1.25in,clip,keepaspectratio]{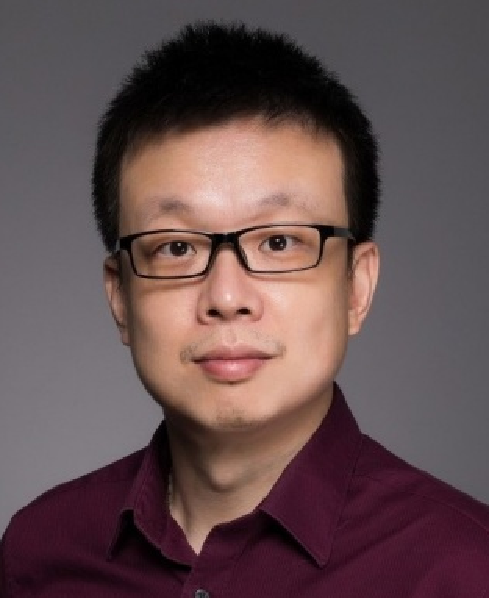}}]{Rui Yang} received the B.Eng. degree in Computer Engineering and the Ph.D. degree in Electrical and Computer Engineering from National University of Singapore in 2008 and 2013 respectively. 

He is currently an Associate Professor in the School of Advanced Technology, Xi’an Jiaotong-Liverpool University, Suzhou, China, and an Honorary Lecturer in the Department of Computer Science, University of Liverpool, Liverpool, United Kingdom. His research interests include machine learning based data analysis and applications. Dr. Yang is currently serving as an Associate Editor for IEEE Transactions on Instrumentation and Measurement, Neurocomputing, and Cognitive Computation.
\end{IEEEbiography}

\begin{IEEEbiography}[{\includegraphics[width=1in,height=1.25in,clip,keepaspectratio]{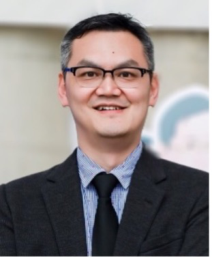}}]{Eng~Gee~Lim}
(M'98-SM'12) received the BEng(Hons) and PhD degrees in Electrical and Electronic Engineering from UK in 1998 and 2002 respectively. Prof. Lim worked for Andrew Ltd, a leading communications systems company in the United Kingdom from 2002 to 2007. Since August 2007, Prof. Lim has been at Xian Jiaotong-Liverpool University, where he was formally the head of EEE department and University Dean of Research and Graduate studies. Now, he is School Dean of Advanced Technology, director of AI university research centre and also professor in department of Communications and Networking. He has published over 200 refereed international journal and conference papers. His research interests are Artificial Intelligence, robotics, AI+ Health care, Future Education, Management in Higher Education, international Standard (ISO/IEC) in Robotics, antennas, RF/microwave engineering, EM measurements/simulations, energy harvesting, power/energy transfer, smart-grid communication; wireless communication networks for smart and green cities. He is a charter engineer and Fellow of both IET and Engineers Australia. In addition, he is also a senior member of IEEE and Senior Fellow of HEA.
\end{IEEEbiography}
\vfill 
\begin{IEEEbiography}[{\includegraphics[width=1in,height=1.25in,clip,keepaspectratio]{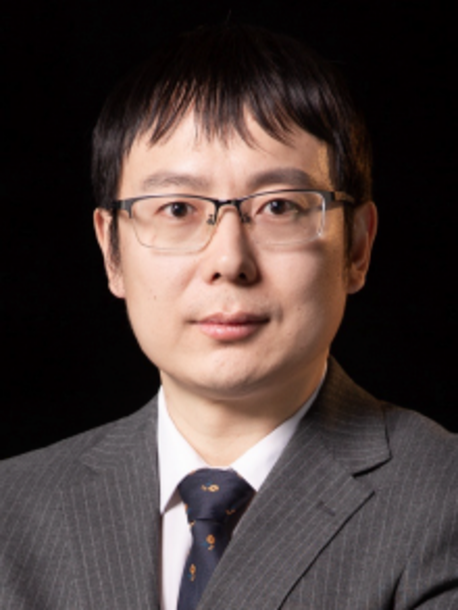}}]{Yutao Yue} (Senior Member, IEEE) is an associate professor at the Artificial Intelligence Thrust and Intelligent Transportation Thrust of Hong Kong University of Science and Technology (Guangzhou). He received his Bachelor’s degree from the University of Science and Technology of China, and Master and PhD degree from Purdue University. He has a dual background in academia and industry, as the team leader of Guangdong Province Introduced Innovation Scientific Research Team, senior scientist of Kuang-Chi Group, and the founder of the Institute of Deep Perception Technology of JITRI. His research interests include multimodal perception fusion, machine consciousness, artificial general intelligence, causal emergence, etc. He has been engaged in scientific research and technology industrialization for over 20 years. He has co-invented 354 granted Chinese patents, 18 USA patents, and 7 EU patents. He has led 6 major research projects with a total funding of nearly 130 million RMB. He has published over 60 papers, advised 13 postdoc research fellows, and received multiple awards including Wu Wenjun Artificial Intelligence Science and Technology Award.
\end{IEEEbiography}
\vfill
\end{document}